\documentclass[a4,11pt]{article}
\pdfoutput=1 

\usepackage{jcappub}
\usepackage[utf8]{inputenc}
\usepackage{graphicx}
\usepackage{color, graphics}
\usepackage{amsmath,amssymb,amsfonts}
\usepackage{verbatim}   
\usepackage{subfigure}  
\usepackage{acronym}
\usepackage{enumerate}   
\usepackage{hyperref}
\usepackage{mathrsfs}
\usepackage{multirow}
 \usepackage{cancel}
 \usepackage{url}
  \usepackage{slashed,afterpage}
 \usepackage{hyperref}
 \usepackage{units}
\usepackage{enumerate}

\def\beq{\begin{equation}\begin{aligned}}
\def\eeq{\end{aligned}\end{equation}}

\begin{document}

\title{\huge Reviving $Z$ and Higgs Mediated Dark Matter Models in Matter Dominated Freeze-out }
\author[1]{Prolay Chanda,}
\author[1,2]{Saleh Hamdan,}
\author[1,3]{and James Unwin}
\affiliation[1]{Department of Physics,  University of Illinois at Chicago, Chicago, IL 60607, USA}
\affiliation[2]{Department of Mathematics,  University of Illinois at Chicago, Chicago, IL 60607, USA}
\affiliation[3]{Department of Physics, University of California, Berkeley \& Theoretical Physics Group, LBNL \& Mathematics Sciences Research Institute, Berkeley, CA 94720, USA}

\abstract{It is quite conceivable that dark matter freeze-out occurred during an early period of matter domination, in which case the evolution and relic abundance differ from standard freeze-out calculations which assume a radiation dominated universe. Here we re-examine the classic  models in which dark matter interactions with the Standard Model are mediated via either the Higgs or $Z$ boson in the context of matter dominated freeze-out. We highlight that while these classic models are largely excluded by searches in the radiation dominated case,  matter dominated freeze-out can relax these limits and thus revive the Higgs and $Z$ portals. Additionally, we discuss the distinctions between matter dominated freeze-out and decoupling during the transition from matter domination to radiation domination, and we comment on the parameter regimes which lead to non-negligible dark matter  production during this transition. 
}

\maketitle


\section{Introduction}

The vast majority of studies of thermal relic dark matter (DM) assume that  DM decouples  whilst the universe is radiation dominated, however this need not be the case. If DM freeze-out occurs whilst the universe is dominated by an energy density which redshifts other than radiation-like, this changes the expansion rate of the universe $H$ and thus will impact the freeze-out dynamics. In particular, previous studies have been made under the assumption that DM freeze-out may have occurred during inflationary reheating \cite{McDonald:1989jd,Chung:1998rq,Giudice:2000ex,Gelmini:2006pw} or kination domination  \cite{Spokoiny:1993kt,Ferreira:1997hj,Salati:2002md,Pallis:2005hm} (whilst the universe was dominated by the kinetic energy of a scalar field) and impacts of non-standard cosmology on DM has recently received renewed attention \cite{Evans:2019jcs,Hamdan:2017psw,Allahverdi:2019jsc,Allahverdi:2018aux,Drees:2017iod,Davoudiasl:2015vba,Delos:2019dyh,Randall:2015xza,Co:2015pka,Dror:2016rxc,DEramo:2017gpl,  DEramo:2017ecx, Garcia:2018wtq,Bernal:2018kcw, Drees:2018dsj, Betancur:2018xtj, Maldonado:2019qmp,Poulin:2019omz,Arias:2019uol,Bramante:2017obj,Bernal:2018ins,Hardy:2018bph,Kaneta:2019zgw,Bernal:2019mhf,Berlin:2016vnh}. 

The prospect that DM freeze-out may have occurred during a period of matter domination was highlighted in  \cite{Hamdan:2017psw}, a possibility observed in passing in  \cite{Kamionkowski:1990ni}. For clarity, we note here that some papers (e.g.~\cite{Allahverdi:2018aux,Drees:2017iod,Drees:2018dsj,Allahverdi:2019jsc}) have used `matter dominated freeze-out' to refer to freeze-in/out during the transition from matter to radiation domination. Here we restrict this term to the case that DM freeze-out  occurs significantly prior to this transition, when entropy injection to the thermal bath can be neglected. Indeed, a  nice feature of matter dominated freeze-out  compared to freeze-out during reheating is that the bath entropy  is conserved and thus the evolution of abundances can be tracked similar to standard  freeze-out calculations. Notably, matter dominated freeze-out  implies a distinct evolution for the DM abundance compared to freeze-out during the transition to radiation domination \cite{McDonald:1989jd,Chung:1998rq,Giudice:2000ex,Gelmini:2006pw}. 

 In \cite{Hamdan:2017psw} matter dominated DM freeze-out was studied in a model independent manner and viable parameter ranges were broadly identified. It was observed that the decay of the matter-like field which leads to the early period of matter domination can dilute the DM, thus allowing for smaller annihilation cross-section or much heavier DM.  Here we continue to explore matter dominated DM freeze-out  in the context of a number of specific models. Specifically, we revisit the classic models of Higgs or $Z$ mediated DM freeze-out in the context of matter dominated freeze-out. We identify scenarios in which these minimal DM scenarios avoid current experimental limits and discuss the prospect for future detection. 

In classic WIMP models DM is a new particle which interacts with the Standard Model only via couplings to either the $Z$ boson or via the Higgs. Electroweak scale  WIMP DM freezing out during radiation domination famously provides the observed relic abundance automatically, and since these couplings and masses are motivated by the Standard Model (and its extensions) this scenario has been affectionately coined the ``WIMP miracle''. Indeed, the classic WIMP picture looks less miraculous with each update from DM searches.  However, under the standard assumption of radiation dominated freeze-out, these classic models are both excluded by searches away from special situations e.g.~\cite{Escudero:2016gzx,Baek:2011aa,Djouadi:2011aa}. These experimental null results have motivated models which ameliorate these experimental limits by either tuning to specific parameter regions, such as resonant annihilation, or generalising the set-up with additional particles and thus introducing more parameter freedom. Here rather than changing the particle physics model, we take an alternative approach,  and explore  whether these simple DM scenarios can be revived if our assumptions on cosmology are different. 

This paper is structured as follows: In Section \ref{s2} we outline the general picture of matter dominated freeze-out, highlighting in particular how it differs from the case of freeze-out during entropy injection which has been studied elsewhere in the literature. In Section \ref{s3} we consider the specific case of the Higgs portal in the most common and minimal realisations involving either scalar or fermion DM. Then in Section \ref{s4} we consider another minimal DM scenario in which the freeze-out DM density is set due to DM annihilations mediated by the Standard Model  $Z$ boson.  In Section \ref{s5} we provide a summary and conclusion, in particular we highlight that if the DM undergoes freeze-out in an early matter dominated era then these classic models can evade current searches and are potentially discoverable.

\section{Matter Dominated Dark Matter Freeze-out}
\label{s2}

 In this section we briefly recap the model independent analysis of matter dominated freeze-out  which appeared in the short paper  \cite{Hamdan:2017psw}, using this opportunity to provide additional relevant details. In particular, we expand on \cite{Hamdan:2017psw} by discussing distinctions with the case in which DM freezes-out whilst $\phi$, the matter-like state which dominates the energy density in the early universe, is decaying. This scenario in which DM freeze-out occurs whilst $\phi$ decays can no longer be neglected is analogous to the case studied in \cite{McDonald:1989jd,Chung:1998rq,Giudice:2000ex,Gelmini:2006pw}. 
 
 We will discuss how the parameter space which gives the correct relic density changes depending on whether DM freeze-out occurs during matter domination, during $\phi$ decays, or after $\phi$ decays (during radiation domination). Finally, we comment on the case that the decaying state $\phi$ has non-negligible branching fraction to DM. In later sections we will re-evaluate the details discussed here in the context of specific models and identify viable regions of parameter space.

\subsection{Model independent look at freeze-out during matter domination}
\label{s2.1}

Consider a population of heavy scalars $\phi$ (in principle $\phi$ could also be a vector boson or fermion), with energy density $\rho_\phi$ and suppose at some critical temperature $T_\star$ that $\rho_\phi$ starts to evolve as matter, i.e.~$\rho_\phi$ scales as $a^{-3}$ in terms of the FRW scale factor. For $T>T_\star$ the contribution of $\phi$ to the energy density is radiation-like, scaling as $a^{-4}$.   The simplest possibility\footnote{More generally if $\phi$ is never in thermal contact with the thermal bath then $T_\star$ has considerable freedom since it denotes the temperature of the Standard Model bath at which $\phi$ becomes matter-like.} is the case that $\phi$ is some heavy species that decouples from the thermal bath then $\phi$  becomes matter-like at $T_\star\sim m_\phi$.

We start  our analysis from the Friedmann equation, given by
	\begin{equation}\label{eq:FE}
		H^2=\frac{8\pi}{3M_{\rm pl}^2}\left[\rho_R+\rho_{\phi}+\rho_{\chi}\right]~,
	\end{equation}
	where  $\rho_R$ and $\rho_{\chi}$ are the energy densities of the Standard Model radiation bath and the DM, respectively.
	For $T_\star>T\gg m_Z,m_\chi$ the Friedmann equation can be rewritten as
	\begin{equation}\label{FEhS}
H^2=H_\star^2\left[\frac{g_* r}{g_*+g_{\chi}}\left(\frac{a_\star}{a}\right)^{4}+(1-r)\left(\frac{a_\star}{a}\right)^{3}+\frac{g_{\chi}r}{g_*+g_{\chi}}\left(\frac{a_\star}{a}\right)^{4}\right]~,
	\end{equation}
in terms of $g_{*}$, the effective number of relativistic degrees of freedom (DoF) of the Standard Model and where $g_\chi$ is DM internal DoF (similarly we define $g_\phi$ to count the internal DoF of $\phi$). Here $H_\star\equiv H(T_\star)$ which can be expressed in terms of $T_\star$ as follows
	\begin{equation}\label{eq:FEf}
		H_\star^2\equiv \frac{8\pi}{3M_{{\rm pl}}^2}\left[\rho_R+\rho_{\phi}+\rho_{\chi}\right]\Big|_{T_\star}=
		\frac{8\pi^3}{90M_{{\rm pl}}^2}\big(g_{*}(T_\star)+g_{\chi}+g_\phi\big)T_\star^{4}~.
	\end{equation}
The quantity $r$ represents the fraction of the energy in radiation at temperature $T_{\star}$, thus $(1-r)$ represents the fraction of energy in the $\phi$ component at $T=T_{\star}$, and $r$ is given by
\beq
	r\equiv\frac{\rho_R+\rho_{\chi}}{\rho_R+\rho_{\chi}+\rho_\phi}\Bigg|_{T=T_\star}~.
	\label{2.4}
\eeq
	
	This form of the Friedmann equation gives a cosmology which is initially radiation dominated for  $r\sim1$, for which $H\propto T^2$. Conversely, matter dominated evolution begins immediately at $T_\star$ for $r\ll 1$ with $H\propto T^{\nicefrac{3}{2}}$. Here $r$ is ratio of energy densities at the initial time and is not a function of time. A good benchmark for $r$ (which we use at various places below) is $r\simeq0.99$ this  corresponds to the case that $\phi$ has $\mathcal{O}(1)$ internal DoF, the DM and Standard Model combined have $\mathcal{O}(100)$ DoF and that all states shared a common temperature in the past which has not significantly deviated by  $T=T_{\star}$.

Assuming that entropy is conserved in the Standard Model radiation, which is a good approximation at times much earlier than the lifetime of $\phi$, then the scale factor is related to the temperature by 
\beq
\label{aeq}
\left(\frac{a_\star}{a}\right)\simeq
\left(\frac{g_*(T)}{g_*(T_\star)}\right)^{\nicefrac{1}{3}} \left(\frac{T}{T_\star}\right) ~.
\eeq
Note, however, that $\phi$ decays eventually lead to entropy non-conservation in the Standard Model bath thus invalidating the relationship of eq.~(\ref{aeq}). Once entropy is no longer conserved in the bath, one instead has a scenario reminiscent to  \cite{McDonald:1989jd,Chung:1998rq,Giudice:2000ex,Gelmini:2006pw} in which freeze-out occurs during a period of entropy injection to the thermal bath, similar to the period of inflationary reheating. We will return to this point shortly and discuss the transition. 	

To proceed we define $x\equiv m_{\chi}/T$ and $x_{\star}\equiv x(T_{\star})$, then neglecting the small DM contribution and using eq.~\eqref{aeq}, the Friedmann equation may be written as
	\begin{equation}\label{eq:FEx}
		H=H_{\star}\sqrt{1-r}\left(\frac{x_{\star}}{x}\right)^{3/2}\left[\frac{r}{1-r}\left(\frac{x_{\star}}{x}\right)+1\right]^{1/2}~.
	\end{equation}
While $\rho_\phi$ evolves as matter, its fractional contribution to the total energy density will grow until it eventually dominates the energy density of the universe,  provided that $\phi$ is sufficiently long-lived. We will discuss the requirement on the $\phi$ lifetime to realise this scenario shortly.  
We define the freeze-out temperature, $T_{f}\equiv m_{\chi}x_{f}^{-1}$ implicitly using the condition:
	\begin{equation}\label{eq:FOcond}
		\Gamma_{\rm ann}(x_{f}):= H(x_{f})~.
	\end{equation}
	Notably, if the DM decouples from the bath whilst the universe is matter dominated this alters the freeze-out dynamics compared to radiation dominated freeze-out, since the Hubble rate scales as $H\propto T^{3/2}$ rather than  $H\propto T^2$.
	
We take the annihilation interaction rate for the DM to be 
	\beq
	\Gamma_{\rm ann}=n_{\chi}^{\rm eq}\langle\sigma_{\rm A}|\vec{v}|\rangle,
	\eeq
	 where $n_{\chi}^{\rm eq}(x)=(2\pi)^{-3/2}gm_{\chi}^{3}x^{-3/2}e^{-x}$
	and we parameterize the thermally averaged annihilation cross-section  as
	\begin{equation}\label{eq:thermAvgPar}
		\langle\sigma_{\rm A}|\vec{v}|\rangle\equiv\hat{\sigma}x^{-n}~.
	\end{equation}
We assume that the DM decouples while non-relativistic (i.e.~$x_f>3$) then from eqns.~\eqref{eq:FEx} \& \eqref{eq:FOcond} we find DM freeze-out occurs at $x_{f}$ for
	\begin{equation}\label{eq:xFO}
		x_{f}=\ln{\left[\frac{gm_{\chi}^{3}\hat{\sigma}}{(2\pi)^{3/2}H_{\star}x_{\star}^{3/2}}(1-r)^{-1/2}\left(x_{f}^{2n}+\frac{rx_{\star}}{1-r}x_{f}^{-1+2n}\right)^{-1/2}\right]}~.
	\end{equation}
	This can be solved approximately via iteration (initially guessing $x_{f}^{(0)}=1$ on the RHS), leading to
	\begin{equation}\label{eq:xFO1}
		x_{f}\approx\ln{\left[\frac{gm_{\chi}^{3}\hat{\sigma}}{(2\pi)^{3/2}H_{\star}x_{\star}^{3/2}}(1-r)^{-1/2}\left(1+\frac{rx_{\star}}{1-r}\right)^{-1/2}\right]}~.
	\end{equation}
	For instance, taking  $r\rightarrow 0$, which implies the evolution starts during matter domination, then  DM freeze-out occurs for
		\begin{equation}\label{eq:xFO1MD}
		x_{f}\big|_{r\rightarrow 0}\simeq\ln{\left[\frac{gm_{\chi}^3\hat{\sigma}}{(2\pi)^{3/2}H_{\star}x_{\star}^{3/2}}\right]}\simeq
		\ln{\left[\frac{3}{4\pi^3}\sqrt{\frac{5}{2}}\frac{g}{\sqrt{g_{*}(T_{\star})}}\frac{m_{\chi}^{3/2}M_{\rm pl}\hat{\sigma}}{\sqrt{T_{\star}}}\right]}~.
	\end{equation}

It is common to define the DM yield $Y\equiv n_\chi/s$ where $s$ is the entropy density $s=\frac{2\pi^2}{45}g_{* S}T^3$ defined in terms of the bath temperature and the effective entropic degrees of freedom $g_{* S}$. One can calculate the yield after freeze-out $Y(x_f)$ assuming it occurs during matter domination and for $n=0$ ($s$-wave) one can evaluate this in the  general $r$ case \cite{Hamdan:2017psw}
	\beq
	Y_{n=0}(x_f)\simeq 
\frac{45}{2\pi^2} \frac{H_\star \sqrt{x_f} x_\star^{5/2}}{
g_{* S}m_\chi^3\hat{\sigma} }\frac{ r}{\sqrt{1-r}}\left(\sqrt{1+\frac{r x_\star}{(1-r)x_f}}-\sqrt{\frac{(1-r)x_f}{r x_\star}}~{\rm sinh}^{-1}\sqrt{\frac{r x_\star}{(1-r)x_f}}\right)^{-1}.
	\label{n=0}\eeq

With the yield $Y(x_f)$ one can normalise this to the critical density and entropy at $x_f$ to compute $\Omega_\chi^{\rm FO}$. However,
following freeze-out the DM abundance is diluted by the decays of the $\phi$ by some factor $\zeta$, during the return to radiation domination, such that the DM relic abundance is 
\beq
\Omega_{\chi}^{\rm relic}=\zeta\Omega_{\chi}^{\rm FO}~.
\eeq Taking  $x_f\simeq23$ (consistent with the reference parameter values we choose below) along with $r\sim0.99$, and assuming an $s$-wave annihilation cross-section of the form $\langle\sigma_{\rm A}|\vec{v}|\rangle\sim \alpha^2/m_\chi^2$ the abundance of DM today is parametrically 
\begin{equation}
\Omega_{\chi}^{\rm relic}h^2\sim 0.1\left(\frac{m_\chi}{10^3~\rm{GeV}}\right)^{\frac{3}{2}}\left(\frac{0.3}{\alpha}\right)^{2} \left(\frac{T_\star}{10^7~\rm{GeV}}\right)^{\frac{1}{2}}\left(\frac{\zeta}{10^{-3}}\right)~.
\label{OMDZ}
    \end{equation}
 
   The quantity $\zeta$ parameterises the change in the entropy of the bath due to $\phi$ decays and can be quantified as follows. Denote by $\Gamma$  the decay width of $\phi$, at the time of the decay of the state $\phi$ then $\Gamma$ should be equal to the Hubble constant $H_{\Gamma}$ and we define a corresponding bath temperature by $\Gamma=H(T_\Gamma$). Assuming the sudden decay approximation for $\phi$ with a decay rate $\Gamma$ then the reheating temperature of the bath following decays $T_{\rm RH}\simeq\sqrt{\Gamma M_{\rm Pl}}$ and the magnitude of the entropy injection $\zeta$ is 
    \beq
\label{eq:ZetaTRH}
&\zeta = \frac{s_{\rm before}}{s_{\rm after}}\simeq\left(\frac{T_{\Gamma}}{T_{\rm RH}}\right)^{3}~,
\eeq
where we neglect any changes in degrees of freedom. It is reasonable to assume the instantaneous decay approximation provided that DM freeze-out occurs well before $H=\Gamma$.

A more useful expression relating $\zeta$ and $T_{\rm RH}$ can be obtain by noting that the duration between the initial condition $a_\star$ and the point $H=\Gamma$ denoted $a_\Gamma$ is given by
\beq\label{Th1c}
  \frac{a_{\star}}{a_{\Gamma}} \approx \frac{1}{(1-r)^{1/3}}\left(\frac{\Gamma}{H_{\star}}\right)^{2/3} ~.
  \eeq
Then we use eq.~(\ref{aeq})  and  that $T_{{\rm RH}}\simeq \sqrt{M_{{\rm Pl}}\Gamma}$ to relate the $T_{\Gamma}$ with $T_{\rm{\rm RH}}$ and $T_{\star}$
\beq
T_{\Gamma}& \simeq \left(\frac{45}{4\pi^{3}g_*(1-r)}\frac{T_{{\rm RH}}^{4}}{T_{\star}}\right)^{1/3}~.
 \eeq
 Substituting $\zeta$ as a function of $T_{\rm RH}$, as in eq.~\eqref{eq:ZetaTRH}, leads to the  following expression for $T_{\rm RH}$
 \beq\label{eq:TRHOmega}
  T_{{\rm RH}}\simeq \frac{4\pi^{3}}{45}  \zeta (1-r) T_{\star} g_*
 ~.
 \eeq
Additionally, it is useful to note that to match the observed relic density for a given freeze-out abundance one requires an entropy injection of order
\begin{equation}\label{eq:zetaChptHiggs}
\zeta=\frac{\Omega_\chi^{\rm relic}h^2}{\Omega_\chi^{\rm FO}h^2}=\frac{\rho_{\rm critical}\Omega_\chi^{\rm relic}h^2}{s_0m_\chi Y(x_f)}~.
\end{equation}

\subsection{The end of the early matter dominated era}

The calculations  of the freeze-out temperature and DM relic abundance presented in Section \ref{s2.1} were based on the assumption of  entropy conservation until all the $\phi$ decay instantaneously at $t\sim\Gamma^{-1}$. We now ascertain when this sudden decay approximation is valid, by comparing with the more natural expectation of an exponential decay law \cite{Scherrer:1984fd}. Notably, an exponential decay law implies a growing source of entropy violation in the thermal bath and at some point the production of radiation due to $\phi$ decays cannot be neglected in the Boltzmann equations.
 In this section we derive when the approximation of entropy conservation breaks down and highlight that there can be considerable parameter space for DM  freeze out to occur whilst entropy injections to the thermal bath from $\phi$ decays are negligible.  In the case that DM decouples from the bath whilst the entropy produced by $\phi$ decays are non-negligible, then DM freeze-out proceeds along similar lines to  Giudice-Kolb-Riotto \cite{Giudice:2000ex}.

The equation governing the decay of a massive particle species $\phi$ can be expressed as
\begin{equation}\label{eq:PhiDecay2}
\frac{d}{dt}\left(\rho_{\phi}a^{3}\right)=-\Gamma\left(\rho_{\phi}a^{3}\right).
\end{equation}
Integrating from $t_\star$, the time when $\phi$ begins evolving like matter ($t_\star\sim H_\star$), to time $t$ gives
\begin{equation}\label{eq:PhiSol}
\rho_{\phi}=\rho_{\phi \star}\left(\frac{a_\star}{a}\right)^{3}e^{-\tau},
\end{equation}
where $\rho_{\phi \star}\equiv \rho_{\phi}(t_\star)$ (similarly, we define $\rho_{R  \star}\equiv \rho_{R }(t_\star)$) and $\tau\equiv\Gamma t$. 
From the Friedmann equations one can show that for an equation of state $\omega$ that the scale factor is related to the time via $a(t)\propto t^{\frac{2}{3}(1+w)^{-1}}$ (where $w=0$ corresponds to matter domination implies  and $w=1/3$ to radiation domination). Then it follows that
\begin{equation}\label{eq:avt2}
\frac{a_\star}{a}=\left(\frac{t_\star}{t}\right)^{\frac{2}{3}(1+w)^{-1}}~.
\end{equation}
Then assuming that $\phi$ decays solely to the radiation sector, the equation governing the radiation energy density can be written as
$\frac{d}{dt}\rho_{R }a^{4}=\Gamma\rho_{\phi}a^{4}.$
Using eq.~\eqref{eq:PhiSol} and eq.~\eqref{eq:avt2} and expanding to zeroth order in $\tau$,
we can  approximate the evolution  as follows
\begin{equation}\label{eq:RDecayApprox}
\frac{d}{dt}\left(\rho_{R }a^{4}\right)\simeq \Gamma\rho_{\phi \star}\left(\frac{t_\star}{t}\right)^{\frac{2}{1+w}}a^{4}\simeq\Gamma\rho_{\phi \star}a_\star^{4}\left(\frac{t_\star}{t}\right)^{-\frac{2/3}{1+w}}~.
\end{equation}
Integrating eq.~\eqref{eq:RDecayApprox} from $t_\star$ to $t$ gives the evolution of the radiation energy density
\begin{equation}\label{eq:RDecaySol}
\rho_R\simeq\rho_{R \star}\left(\frac{a_\star}{a}\right)^4+\rho_{\phi\star}\left(\frac{a_\star}{a}\right)^4\frac{\tau}{(\nicefrac{2}{3})(1+w)^{-1}+1}\left[\left(\frac{\tau_\star}{\tau}\right)^{-\frac{2/3}{1+w}}-\frac{\tau_\star}{\tau}\right]~,
\end{equation}
where $\tau_\star\equiv \Gamma t_\star$. The first term on the RHS of eq.~\eqref{eq:RDecaySol} signifies the ``old'' radiation present at $T=T_\star$, and the latter term is the ``new'' radiation  corresponding to the decay products of $\phi$. Furthermore, using that $\rho_{R }\simeq r\rho_\star$ and $\rho_{\phi\star}\simeq (1-r)\rho_\star$, where we neglect the small $\rho_{\chi}$ component, and defining 
\beq
\mathcal{T}\equiv\frac{\tau_\star}{\tau}=\frac{t_\star}{t}~,
\eeq
we can rewrite eq.~\eqref{eq:RDecaySol} as follows
\begin{equation}\label{eq:RDecaySolT}
\frac{\rho_R}{\rho_\star}\simeq r\mathcal{T}^{\frac{8/3}{1+w}}+\frac{(1-r)\tau_\star}{(\nicefrac{2}{3})(1+w)^{-1}+1}\left(\mathcal{T}^{\frac{2}{1+w}-1}-\mathcal{T}^{\frac{8/3}{1+w}}\right)~.
\end{equation}

\begin{figure}[t!]
\centering
	\advance\leftskip-12mm
	\includegraphics[width=0.55\textwidth]{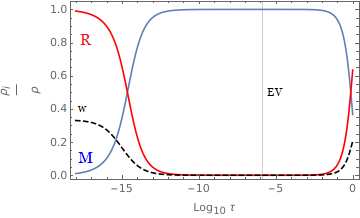}~
	\includegraphics[width=0.47\textwidth]{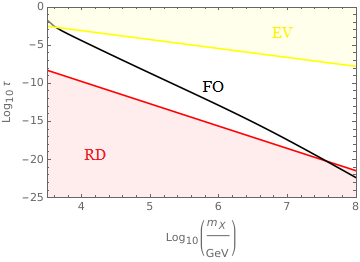}
	\vspace{-3mm}
\caption{
	Left: Evolution of the fraction of energy density in matter (blue) and radiation (red), for $T_{\rm RH}=10$ MeV, $T_\star=10^7$ GeV and $r=0.99$. The $w$ curve (dashed) parameterizes the change from radiation domination ($w=1/3$) to matter domination ($w=0$). The first intersection of blue and red line signifies the onset of matter domination and the vertical black line shows the point $T=T_{\rm EV}$  at which entropy production in the bath is non-negligible.  
	Right: Orderings of events in time parameter $\tau$ as a function of DM mass  for $T_\star=10^7$ GeV, and $r=0.99$ with $T_{\rm RH}$ fixed to reproduce observed DM relic density. The black line indicates the time of freeze-out, the radiation domination era shaded in red, the period in which entropy violation is considerable is shaded in yellow, and matter domination freeze-out occurs in the white region.}
	\label{fig:EV}
\end{figure}

In their classic paper  \cite{Scherrer:1984fd}, Scherrer and Turner showed that the entropy in the thermal bath remains roughly constant until the old and new radiation are comparable (that is the first and second terms of the RHS of eq.~\eqref{eq:RDecaySolT}, respectively). Thus the point at which the assumption of entropy conservation is violated $\mathcal{T}_{\rm EV}$ can be defined implicitly by
\begin{equation}\label{eq:TentDef}
r\mathcal{T}_{\rm EV}^{\frac{8/3}{1+w}}:=\frac{(1-r)\tau_\star}{(\nicefrac{2}{3})(1+w)^{-1}+1}\left(\mathcal{T}_{\rm EV}^{\frac{2}{1+w}-1}-\mathcal{T}_{\rm EV}^{\frac{8/3}{1+w}}\right)~.
\end{equation}
Solving eq.~\eqref{eq:TentDef} for $\mathcal{T}_{\rm EV}$, and writing $v\equiv(\nicefrac{2}{3})(1+w)^{-1}+1$ gives
\begin{equation}\label{eq:TEV}
\mathcal{T}_{\rm EV}=\left(1+\left(\frac{r}{1-r}\right)\frac{v}{\tau_\star}\right)^{-1/v}~.
\end{equation}

Since entropy remains roughly constant until $t=t_{\rm EV}\equiv t_\star/\mathcal{T}_{\rm EV}$, it follows that $T\propto a^{-1}$ prior to $t=t_{\rm EV}$ and $T\propto a^{-1}$ becomes invalid once  entropy conservation is strongly violated. Evolving from $T_\star$  and using eq.~(\ref{eq:TEV}) we obtain the following expression for the point of entropy conservation violation 
\begin{equation}\label{eq:TempEnt2}
T_{\rm EV}\equiv T(t_{\rm EV})\simeq T_\star \left(\frac{t_\star}{t_{\rm EV}}\right)^{v-1}
 \simeq T_\star \left(1+\left(\frac{r}{1-r}\right)\frac{v}{\tau_\star}\right)^{(1-v)/v}.
\end{equation}
This point $T_{\rm EV}$ marks the threshold between the matter dominated period and the particle decay era. Thus for $T\gtrsim T_{\rm EV}$ we have that $T\propto a^{-1}$ and $H\propto T^{3/2}$, whereas for $T\lesssim T_{\rm EV}$ then $T\propto a^{-3/8}$ and $H\propto T^{4}$ \cite{Scherrer:1984fd}.
For freeze-out to occur during matter domination (which is the focus of this work), rather than during particle decays (as in \cite{Giudice:2000ex}), it is required that:
\begin{equation}\label{eq:TEVcons}
T_f\gtrsim T_{\rm EV}~.
\end{equation}

\newpage

Figure \ref{fig:EV} (left) shows the evolution of the energy densities for fixed $T_\star=10^7$ GeV taking $r=0.99$ and $T_{\rm RH}=10$ MeV. For $w$ we take an interpolation between $w=1/3$ and $w=0$ weighted linearly according to the relative fraction of energy that radiation and matter constitute. It can be seen that it is possible to have a significant amount of time between when matter domination ensues (as indicated by the intersection of the matter and radiation curves) until the point ``EV'' when entropy violations become considerable. 

For Figure \ref{fig:EV} (right) we do not fix $T_{\rm RH}$  (the other parameters are fixed the same), rather  $T_{\rm RH}$ is adjusted such that the observed DM relic density is reproduced. The pair of plots in Figure \ref{fig:EV} indicate the viability of DM freeze-out in the matter dominated regime.  The distinction between our work and previous related studies \cite{McDonald:1989jd,Giudice:2000ex,Gelmini:2006pw,Chung:1998rq} is most clearly seen in the right panel which demonstrates that the time of freeze-out (black curve) can readily occur after matter domination begins and prior to when entropy injection is non-negligible.  

\subsection{Production of dark matter in the transition to radiation domination}
\label{s2.3}

Throughout this work we assume that the DM relic density is set during matter domination and that there is no subsequent production of DM during the transition to radiation domination. This implies that the DM does not couple to $\phi$ or has only sufficiently small coupling such that the production is subleading to the freeze-out abundance. As noted above, the converse case of DM produced during the transition has been studied in e.g.~\cite{Chung:1998rq,McDonald:1989jd,Giudice:2000ex,Gelmini:2006pw}. However, since the DM couples to Standard Model there are loop induced processes which couple $\phi$ to the DM and in this section we start to quantify over which parts of parameter space this implies substantial DM production during $\phi$ decays.

Denote by $\Omega_{\chi, {\rm decay}}$  the contribution to the late time DM abundance due to $\phi$ decays to DM  and by $\Omega_{\chi, {\rm MDFO}}$ the contribution coming from matter dominated DM freeze-out (and subsequent dilution). The condition that matter dominated DM freeze-out sets the observed DM  density today can be surmised as $\Omega_{\chi, {\rm decay}}<\Omega_{\chi, {\rm MDFO}}$. However, we shall suppose that  after dilution the matter dominated DM freeze-out abundance matches the observed relic density $\Omega_{\chi, {\rm MDFO}}\approx\Omega_{\chi, {\rm relic}}h^{2}\approx0.1$ and thus the condition that $\phi$ decays to DM can be neglected can be restated as   $\Omega_{\chi, {\rm decay}}h^{2}\ll0.1$. 

The Boltzmann equation for the DM number density $n_{\chi}$ is of the form
\beq\label{eq:DMnumberDensity0}
&\frac{dn_{\chi}}{dt} + 3Hn_{\chi} = 2\Gamma\frac{\rho_{\phi}}{m_{\phi}}\mathcal{B}_{\rm DM}~,
 \eeq
where $\mathcal{B}_{\rm DM}$ branching ratio of $\phi$ to DM pairs.
 Then the contribution $\Omega_{\chi, {\rm decay}}$ due to $\phi$ decay to DM assuming a branching ratio to DM of $\mathcal{B}_{\rm DM}$  is given by \cite{Kaneta:2019zgw} (rederived in Appendix \ref{DMdensityPhiDecay})
\beq
\frac{\Omega_{\chi, {\rm decay}}h^{2}}{\Omega_{\chi,{\rm MDFO}}h^{2}} &\simeq 0.1\times \left(\frac{\mathcal{B}_{\rm DM}}{10^{-8}}\right)\left(\frac{10~{\rm TeV}}{m_{\phi}}\right)
\left(\frac{T_{{\rm RH}}}{1~{\rm GeV}}\right)\left(\frac{m_{\chi}}{15~{\rm GeV}}\right)~.
\label{bran}
\eeq
The selected parameters imply a limiting case in which production accounts for 10\% of DM. Larger $m_\phi$, or smaller $T_{\rm RH}$ or $m_\chi$,  significantly suppress $\Omega_{\chi, {\rm decay}}$ and give  $\Omega_{\chi, {\rm decay}}h^{2}\ll0.1$. 

Loop induced decays  to DM via Standard Model states lead to a non-zero branching fraction to DM, but the form of the loop induced branching ratio to DM will depend critically on the portal operator. Thus we will evaluate $\mathcal{B}_{\rm DM}$ for each model we subsequently explore in turn. As we will discuss for each of the portals in what follows, the branching ratio can typically be sufficiently small that one can arrange for loop induced production of DM due to $\phi$ decays to be negligible in most of the models we study.

\section{Matter Dominated Freeze-out via the Higgs Portal}
\label{s3}

We next apply our previous considerations to a number of specific DM models,  starting with the Higgs Portal. The Higgs boson provides a promising means to link DM to the Standard Model \cite{Silveira:1985rk,McDonald:1993ex,Burgess:2000yq,Patt:2006fw}. The precise relations of the Higgs Portal depend on the nature of the DM quantum numbers. The best motivated Higgs portal models are those involving scalar or fermion DM; the former is aesthetic as it is highly minimal, requiring the addition of only a single new state, whereas the fermion case typically requires both a new fermion and some additional mediator state, but has the advantage that the DM can be light without running into naturalness concerns.

 \subsection{Scalar dark matter via the Higgs portal}

We  first consider the case of scalar DM\footnote{In the interim between the initial letter \cite{Hamdan:2017psw} and thesis of SH \cite{Hamdan:2018fqj} and the completion of this work, two articles appeared \cite{Hardy:2018bph,Bernal:2018ins} which study aspects of matter dominated freeze-out via the scalar Higgs portal.} $\chi$, with a $\mathbb{Z}_2$ invariant Lagrangian  \cite{Silveira:1985rk,McDonald:1993ex,Burgess:2000yq,Patt:2006fw}
 \beq \label{Hlag}
 \mathcal{L} = \mathcal{L}_{\text{SM}}+\frac{1}{2}\partial_{\mu}\chi\partial^{\mu}\chi -\frac{1}{4!}\lambda\chi^{4}-\frac{1}{2}\mu_{\chi}^{2}\chi^{2}-\frac{c_{\chi}}{2}\kappa \chi^{2}H^{\dagger}H~,
 \eeq
where $H$ is the Standard Model Higgs, $\mathcal{L}_{\rm SM}$ is the Standard Model Lagrangian, $\kappa$ describes coupling of the  scalar DM with the Higgs boson, $\mu_{\chi}$ is the bare mass of the DM, and the constants $c_{\chi} = 2$  for a complex scalar and $c_\chi=1$ for a real scalar.
While $\lambda$ is often reserved for the  Higgs quartic couplings, since we won't make reference to this quantity, we use $\lambda$ instead for the $\chi$ quartic coupling and importantly this is unrelated to the Higgs coupling. 

After electroweak symmetry breaking the Higgs boson acquires a vacuum expectation value (VEV) $\langle H\rangle=v_0$ and  the Lagrangian may be expanded  around this VEV giving
\begin{equation}\label{eq:scalarLagrangeExpand}
 \mathcal{L} = \mathcal{L}_{{\rm SM}}+\frac{1}{2}\partial_{\mu}\chi\partial^{\mu}\chi -\frac{1}{4!}\lambda\chi^{4}-\frac{1}{2}m_{\chi}^{2}\chi^{2}-\frac{c_{\chi}}{4}\kappa \chi^{2}h^{2}-\frac{c_{\chi}}{2}\kappa\chi^{2}v_{0}h~.
\end{equation}
The DM $\chi$ receives contribution in mass from the bare mass term and the cross-coupling term and can be expressed as $ m_{\chi} = \sqrt{\mu_{\chi}^{2}+\frac{c_{\chi}}{2}\kappa v_{0}^{2}}~.$ In what follows we shall consider DM with mass in excess of 10 GeV and $\kappa\lesssim0.03$ in which case  $m_{\chi} \simeq \mu_{\chi}$ and is a free parameter.

The three main annihilation routes for scalar DM  via the Higgs portal are (i).~DM annihilations to Standard Model fermions $\chi\chi\rightarrow f\bar{f}$, (ii).~DM to Standard Model vector bosons, $\chi\chi\rightarrow V\bar{V}$, (iii).~DM to Higgs bosons, $\chi\chi\rightarrow hh$. The corresponding partial cross-sections in terms of the Mandelstam variable $s$ are
 \beq
 \sigma_{\chi\chi\to f\bar{f}} &= \frac{N_{c}c_{\chi}^{2}\kappa^{2}m_{f}^{2}}{8\pi s}\frac{\sqrt{1-\frac{4m_{f}^{2}}{s}}}{\sqrt{1-\frac{4m_{\chi}^{2}}{s}}}\frac{s-4m_{f}^{2}}{(s-m_{h}^{2})^{2}+m_{h}^{2}\Gamma_{h}^{2}}\\
 \sigma_{\chi\chi\to V\bar{V}} & =\frac{\delta_{V}c_{\chi}^{2}\kappa^{2}m_{V}^{4}}{4\pi s}\frac{\sqrt{1-\frac{4m_{V}^{2}}{s}}}{\sqrt{1-\frac{4m_{\chi}^{2}}{s}}}\frac{\left(2+\frac{\left(s-2m_{V}^{2}\right)^{2}}{4m_{V}^{4}}\right)}{\left[(s-m_{h}^{2})^{2}+(m_{h}\Gamma_{h})^{2}\right]}\\
 \sigma_{\chi\chi\to hh} &= \frac{\kappa^{2}c_{\chi}^{2}}{32\pi s}\frac{\sqrt{1-\frac{4m_{h}^{2}}{s}}}{\sqrt{1-\frac{4m_{\chi}^{2}}{s}}} ~,
 \eeq
where $N_{c}$ is the number of colours, $V=Z,W^\pm$ with $\delta_{Z} = \frac{1}{2}$ and $\delta_{W^{+},W^{-}}=1$,  $m_h$ is the Higgs boson mass and the Higgs width is $\Gamma_h^{\mathrm{SM}}\approx13~\mathrm{MeV}$ \cite{Tanabashi:2018oca}.

\begin{figure}[t!]
		\advance\leftskip-25mm
	\centerline{
		\includegraphics[width=0.53\textwidth]{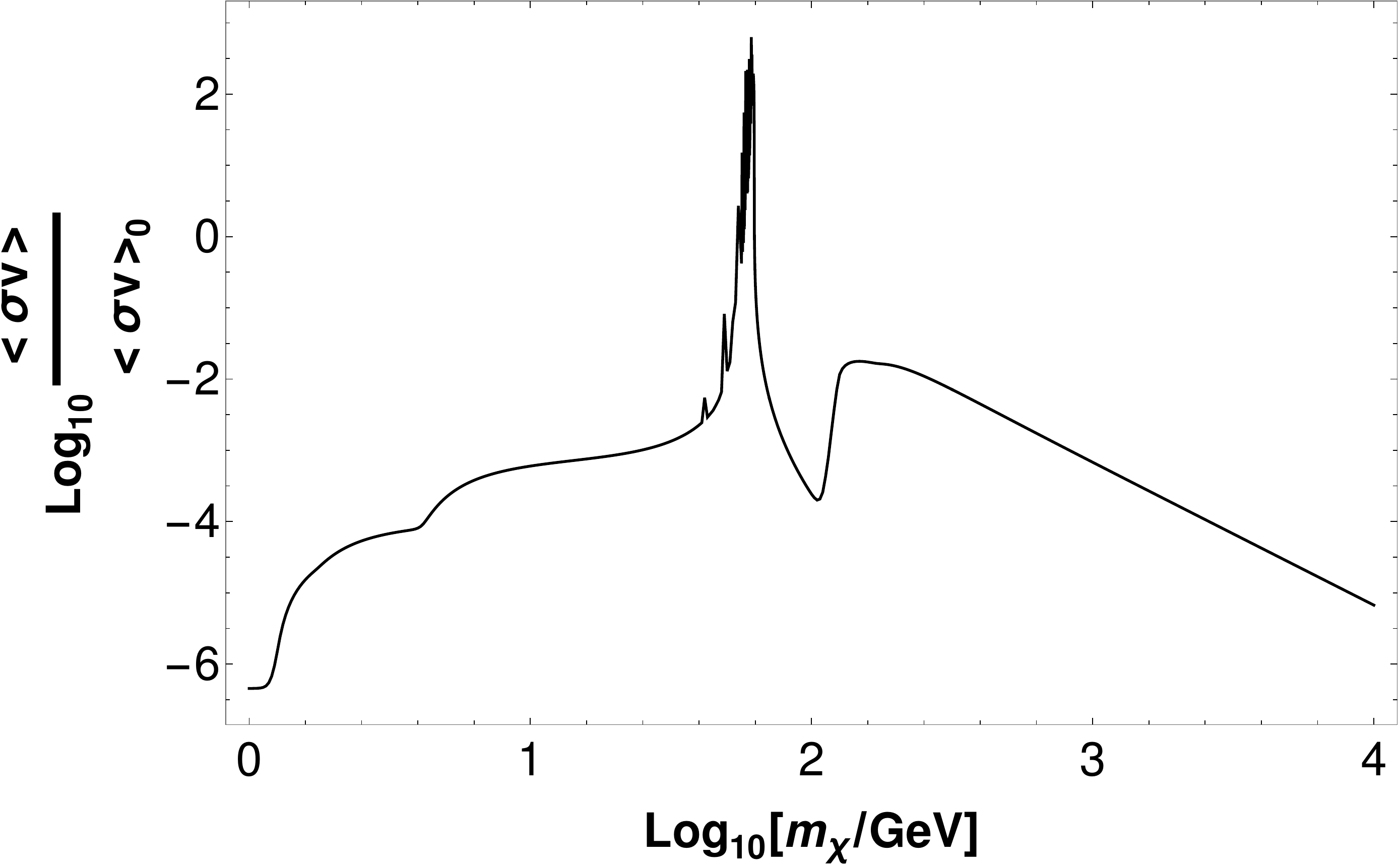}
	\includegraphics[width=0.53\textwidth]{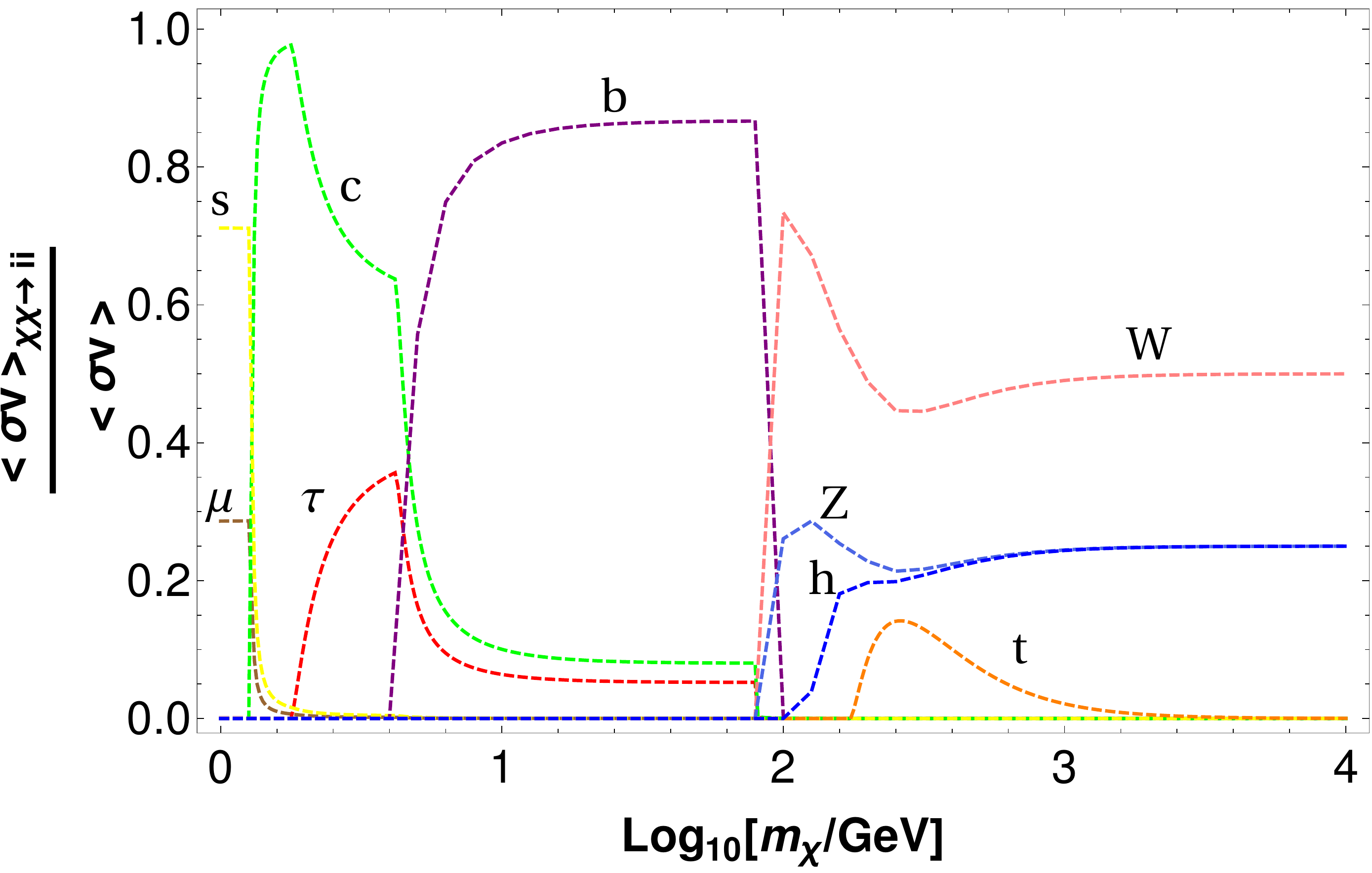}}
	\caption{Left: The total Higgs Portal annihilation cross-section for the complex scalar DM in units of $\langle\sigma v\rangle_0=3\times10^{-26}$cm$^3/$s, for a fixed value of the coupling $\kappa=0.01$ of the mixed quartic $|H|^2|\chi|^2$. Right: The fraction of the total annihilation cross-section at freeze-out going to various Standard Model final states.}
	\label{fig:TotalCSscalarHportal}
\end{figure}

With the leading contributions to the annihilation cross-section, we now calculate the relic density of DM after freeze-out.  The Boltzmann equation can be expressed in terms of the thermally-averaged cross-section $\langle\sigma v\rangle$ which is a product of the  annihilation cross-section and relative DM velocity averaged over the equilibrium distribution functions  \cite{Gondolo:1990dk}
\begin{equation}\label{eq:thermAvgGond}
\langle\sigma v\rangle=\frac{1}{8m_\chi^4TK_2^2(x)}\int^\infty_{4m_\chi^2}\sigma (s-4m_\chi^2)\sqrt{s}K_1\left(\sqrt{s}/T\right)ds~.
\end{equation}
 Thermally averaging term-by-term gives $\langle\sigma v\rangle=\sigma_0+\sigma_1v^{2}+\cdots$ In the non-relativistic limit, expanding the partial thermally averaged cross-sections $\langle\sigma v\rangle_i$ to order $v^{2}$ gives
 \beq\label{eq:sDMannH}
 \langle\sigma v\rangle_{\chi\chi\to f\bar{f}} &=
 \frac{N_{c}c_{\chi}^{2}\kappa^{2}r_{f}^{2} }{16\pi m_{\chi}^{2}}\frac{\left(1-4r_{f}^{2}\right)^{3/2}}{\left(1-r_{h}^{2}\right)^{2}+r_{\Gamma}^{2}}\left[1
+\frac{v^2}{2}\frac{7r_{f}^{2}+r_{h}^{2}(1-10r_{f}^{2})+3r_{f}^{2}(r_{h}^{4}+r_{\Gamma}^{2})-1}{\left(1-4r_{f}^{2}\right)\left[\left(1-r_{h}^{2}\right)^{2}+r_{\Gamma}^{2}\right]}\right]\\[5pt]
 \langle\sigma v\rangle_{\chi\chi\to V\bar{V}} &=
 \frac{\delta_{V}c_{\chi}^{2}\kappa^{2}}{32\pi  m_{\chi}^{2}}\sqrt{1-4r_{V}^{2}}
 \frac{(1-4r_{V}^{2}+12r_{V}^{4})}{\left(1-r_{h}^{2}\right)^{2}+r_{\Gamma}^{2}} \left[ 1
 +\frac{v^2}{4} \frac{F-1}{\left(1-4r_{V}^{2}+12r_{V}^{4}\right)\left[(1-r_{h}^{2})^{2}+r_{\Gamma}^{2}\right]}\right]\\[5pt]
  \langle\sigma v\rangle_{\chi\chi\to hh}  &=\frac{\kappa^{2}c_{\chi}^{2}}{64\pi m_{\chi}^{2}}\sqrt{1-4r_{h}^{2}}\left[1
 +\frac{v^2}{4}\frac{6r_{h}^{2}-1}{1-4r_{h}^{2}}\right]~,
 \eeq
in terms of the ratios
 $r_{i} = m_{i}/(2m_{\chi})$  and  $r_{\Gamma}=m_{h}\Gamma_{h}/(4m_{\chi}^{2})$,
 and where $F$ is the following combination of these various factors
 \beq
 F=14r_{V}^{2}-76r_{V}^{4}+168r_{V}^{6}-r_{h}^{2}(12r_{V}^{2}-96r_{V}^{4}+240r_{V}^{6})+(r_{h}^{4}+r_{\Gamma}^{2})(1-2r_{V}^{2}-20r_{V}^{4}-72r_{V}^{6}).
\notag
 \eeq

Figure \ref{fig:TotalCSscalarHportal} (left) shows the total DM annihilation cross-section to Standard Model states via the Higgs portal (in units of $\langle\sigma v\rangle_0=3\times 10^{-26}$cm$^3/$s being the conventional DM thermal cross-section) for a fixed value $\kappa=0.01$ of the coupling constant of mixed quartic $|H|^2|\chi|^2$. 
Additionally, in Figure \ref{fig:TotalCSscalarHportal} (right) we present the fraction of the total annihilation cross-section at freeze-out going to various Standard Model final states which depends on $m_\chi$ but not $\kappa$. Our results  are in good agreement with previous studies of the Higgs portal e.g.~\cite{Cline:2012hg,Cline:2013gha,Feng:2014vea,Escudero:2016gzx}.  

\begin{figure}[t!]
\begin{center}
{\bf Higgs Portal for Scalar Dark Matter: Consistency Conditions}
\end{center}
	\centerline{ 
	\includegraphics[width=0.52\textwidth]{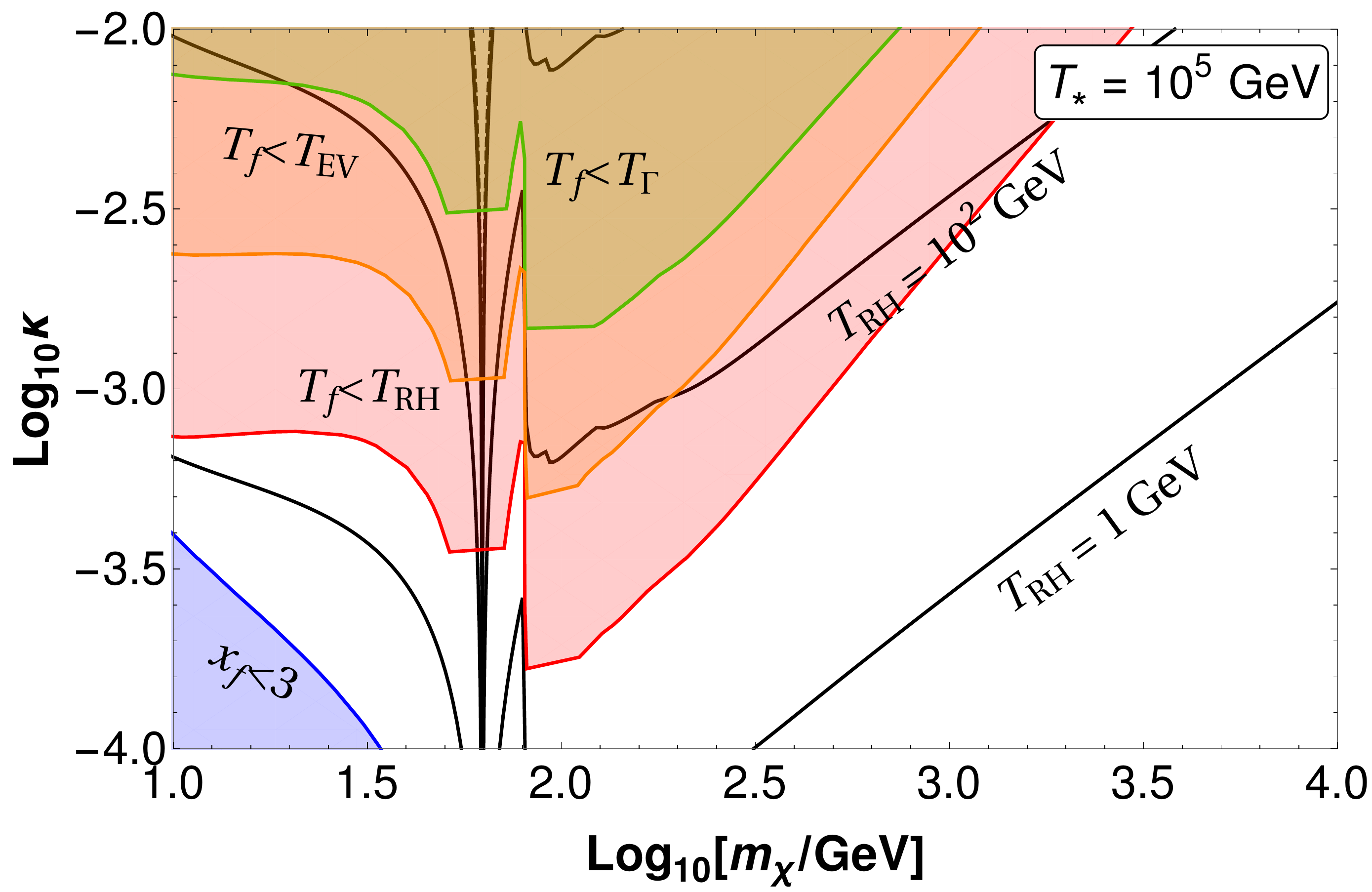}
	\includegraphics[width=0.52\textwidth]{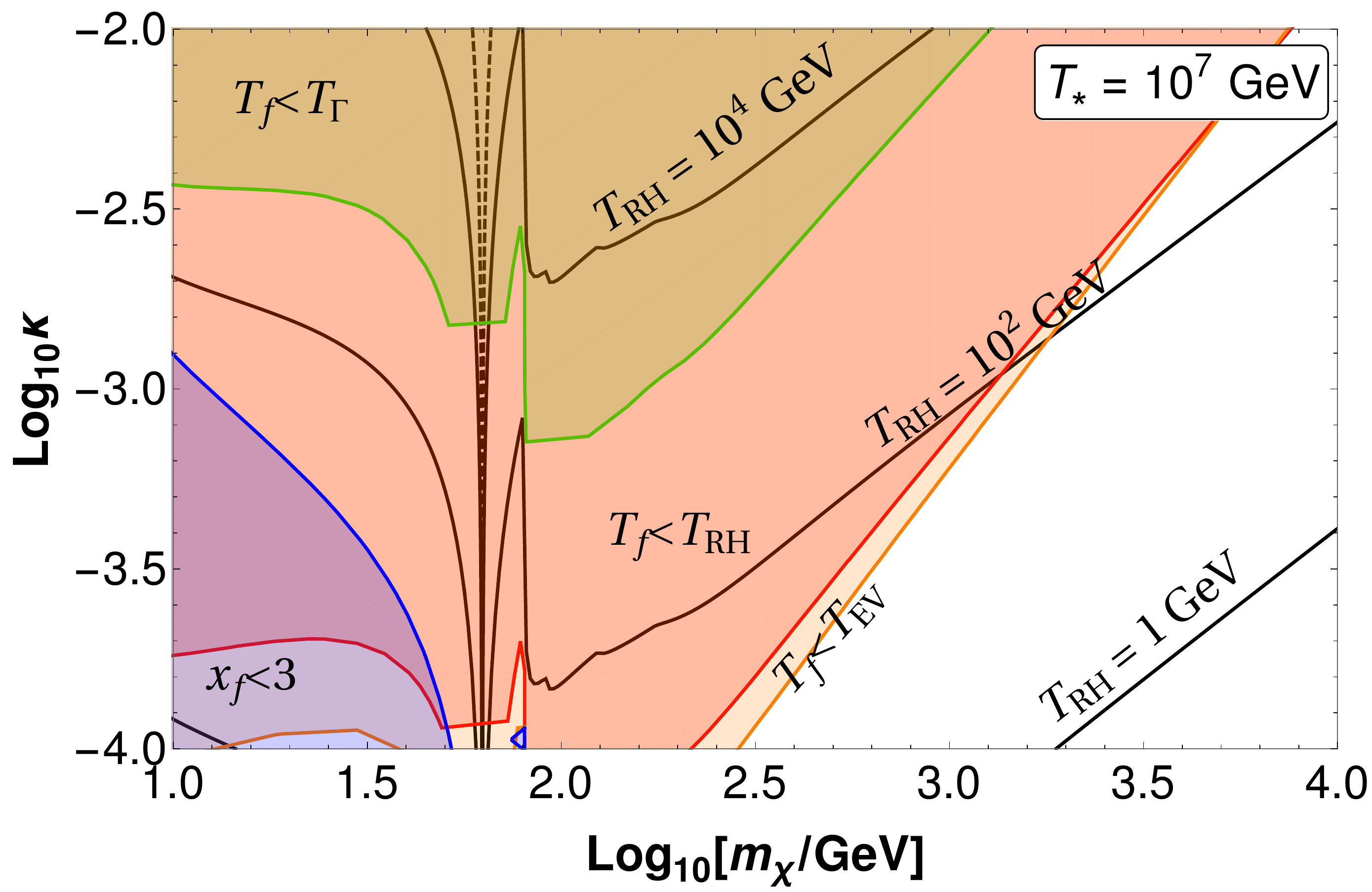}}
\vspace{-3mm}
	\caption{The above plots examine the case of complex scalar DM annihilating through the Higgs portal  for the parameter values $r = 0.99$ and $T_{\star} = 10^{5}$ GeV (left) and $10^{7}$ GeV (right). The solid black curves show contours of the reheat temperature ($T_{\rm RH}$) which give the observed DM relic density. The dashed black line shows the analogous case of radiation dominated DM freeze-out (without an entropy injection). The shaded regions signify parameter values for which the consistency conditions fail: $\phi$ decay before freeze out $T_{f}<T_{\Gamma}$ (green),  decays of $\phi$ cannot be neglected $T_f\ll T_{\rm EV}, T_{\rm RH}$ (red, orange), and DM decouples relativistically with $x_f<3$ (blue). Observe that viable parameter space remains with the observed DM relic density.}
	\label{fig:kContour-ScalarDM}
\end{figure}

With the thermally averaged cross-sections above we can solve the Boltzmann equations and determine the freeze-out abundance. Similar to the model independent  analysis of Section \ref{s2} the freeze-out abundance will be diluted by a factor $\zeta$ during the transition from matter to radiation domination, such that $\Omega_{\rm DM}^{\rm relic}h^2=\zeta\Omega_\chi^{\rm FO}h^2$. Note that $\zeta$ can be re-expressed in terms of $T_{\rm RH}$ via eq.~\eqref{eq:TRHOmega}. In Figure \ref{fig:kContour-ScalarDM} we display contours of the $T_{\rm RH}$ which give the correct relic density for scalar DM freezing out via the Higgs portal (black solid curves). The plot is overlaid with cosmological consistency constraints which we discuss in the next section.

Furthermore, we highlight that for a specific DM mass range one can derive reasonably succinct analytic expressions. For instance, consider $10~{\rm GeV}\lesssim m_{\chi}\lesssim  80$ GeV in which case the DM mainly annihilates to $\overline{b}b$ and as a result if freeze-out occurs during matter domination then the DM decouples at a temperature
\beq
x_{f} = \ln\left(\sqrt{\frac{45}{28\pi^{8}}}\frac{N_{c}c_{\chi}^{2}\kappa^{2}gM_{{\rm Pl}}}{64g_*^{1/2}T_{\star}^{1/2}}(1-r)^{-1/2}\frac{m_{b}^{2}}{m_{\chi}^{5/2}}\frac{\left(1-\frac{m_{b}^{2}}{m_{\chi}^{2}}\right)^{3/2}}{\left[\left(1-\frac{m_{h}^{2}}{4m_{\chi}^{2}}\right)^{2}+\left(\frac{m_{h}\Gamma_{h}}{4m_{\chi}^{2}}\right)^{2}\right]}\right)~.
\eeq
Given this value of $x_f$ the resulting DM relic density can be expressed as
 \beq
\Omega_{{\rm DM}}^{{\rm Relic}}h^{2} &\simeq 2\times 10^{11}~{\rm GeV}^{-1} \zeta
\frac{ \left(1-r\right)^{1/2} g_*^{-1/2}}{N_{c}c_{\chi}^{2}\kappa^{2}M_{{\rm Pl}}}\frac{m_{\chi}^{4}}{m_{b}^{2}}\frac{x_{f}^{3/2}}{x_{\star}}\frac{\left[\left(1-\frac{m_{h}^{2}}{4m_{\chi}^{2}}\right)^{2}+\left(\frac{m_{h}\Gamma_{h}}{4m_{\chi}^{2}}\right)^{2}\right]}{\left(1-\frac{m_{b}^{2}}{4m_{\chi}^{2}}\right)^{3/2}}~.
\eeq
The explicit GeV${}^{-1}$ and numerical factors appear due to the substitution of the critical density and entropy density today.

\subsection{Constraints on the scalar Higgs portal}

The freeze out calculations above are based on a number of assumption, in particular about the hierarchies  of energy scales and orderings of events, in this section we first discuss the cosmological consistency conditions which must be satisfied for our analysis to hold. Specifically, we have introduced a number of temperature scales: $T_{f}$ at which  DM freeze-out occurs, $T_{\Gamma}$ at which $\phi$ decays, $T_{\rm RH}$ the reheat temperature and $T_{\rm EV}$ at which $\phi$ decays cannot be neglected in the Boltzmann equations.  Let us also denote by $T_{\rm MD}$ the temperature at which matter domination begins. 

With the above scales defined we can discuss the consistency conditions which are required to be satisfied to be in the regime of matter dominated freeze-out, these are:
\begin{itemize}
	\vspace{1mm}
	\item Freeze-out of dark matter during matter domination ($T_f< T_{\rm MD}\leq T_{\star}$).
	\vspace{1mm}
	\item Sufficiently high reheat temperature from $\phi$ decays ($T_{{\rm RH}}>T_{\rm BBN}\sim 10$ MeV).
	\vspace{1mm}
	\item Freeze-out of dark matter before majority decays of the $\phi$  states ($T_f\gg T_{\Gamma}$). 
	\vspace{1mm}
	\item Freeze-out of dark matter while decays of $\phi$ can be neglected ($T_f\gg T_{\rm EV}, T_{\rm RH}$). 
	\vspace{1mm}
	\item Freeze-out of dark matter while non-relativistic  ($T_f<m_\chi/3$). 
	\vspace{1mm}
\end{itemize}

We apply these constraints to the scalar Higgs portal in Figure \ref{fig:kContour-ScalarDM} for two values of the critical temperature $T_\star$. For $T_{f}<T_{\Gamma}$  DM decouples after decays and during radiation domination, thus our analysis is inappropriate, this case is shown as the green shaded region in Figure \ref{fig:kContour-ScalarDM}. The red and orange regions exclude the parameter space which predicts that decays of $\phi$ cannot be neglected $T_f\not\gg T_{\rm EV},T_{\rm RH}$.  There is also a region in which $T_f>m_\chi/3$ in which case the DM decouples relativistically and is excluded by classic `Hot DM' bounds. Finally, there is the requirement that the reheat temperature is greater than a few MeV \cite{Sarkar:1995dd,Hannestad:2004px,deSalas:2015glj} in order to reproduce standard big bang nucleosynthesis (BBN), and we will impose the slightly more conservative condition $T_{{\rm RH}} > 10 $ MeV. While this BBN constraint does not show up on the plots of  Figure \ref{fig:kContour-ScalarDM}, it will appear in subsequent plots.

Note, the orange region indicates the region in which the inequality of  eq.~\eqref{eq:TEVcons}  is no longer satisfied, implying that entropy in the thermal bath is no longer conserved, and thus a breakdown in the validity of the calculation. Let us re-emphasise that $T_f<T_{\rm EV},T_{\rm RH},T_\Gamma$ does not immediately imply an exclusion, but rather indicates that a separate calculation is needed assuming freeze-out during entropy injection \cite{McDonald:1989jd,Chung:1998rq,Giudice:2000ex,Gelmini:2006pw}  or during radiation domination (possibly with an entropy dump \cite{Berlin:2016vnh,Bramante:2017obj}). These alternative cosmologies may also potentially yield the correct relic density while evading constraints, and have been discussed in the context of the scalar Higgs portal in \cite{Hardy:2018bph,Bernal:2018ins}. 

\begin{figure}[t!]
\begin{center}
{\bf  Higgs Portal for Scalar Dark Matter: Experimental  Limits}
\end{center}
\centerline{	\includegraphics[width=0.52\textwidth]{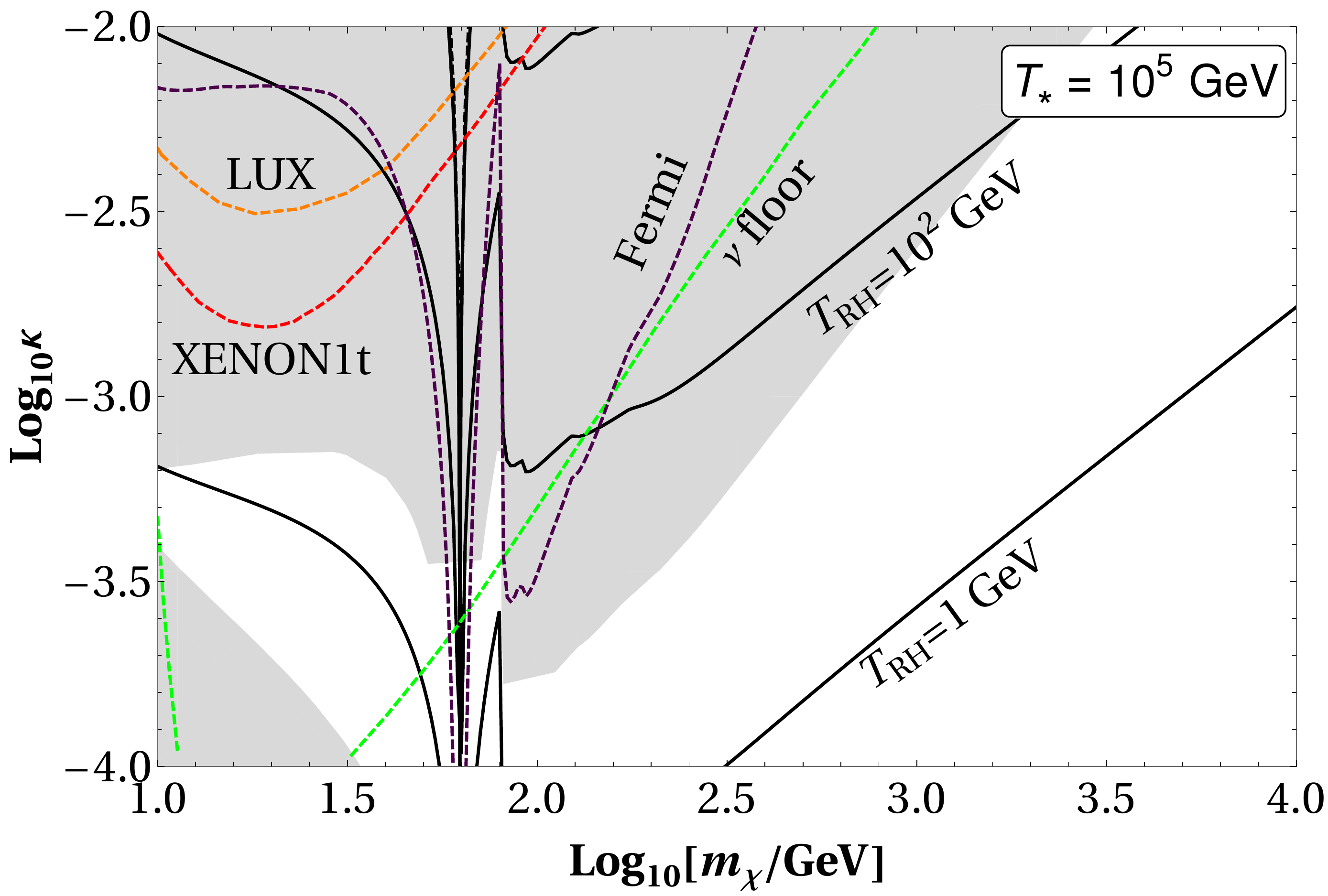}
	\includegraphics[width=0.52\textwidth]{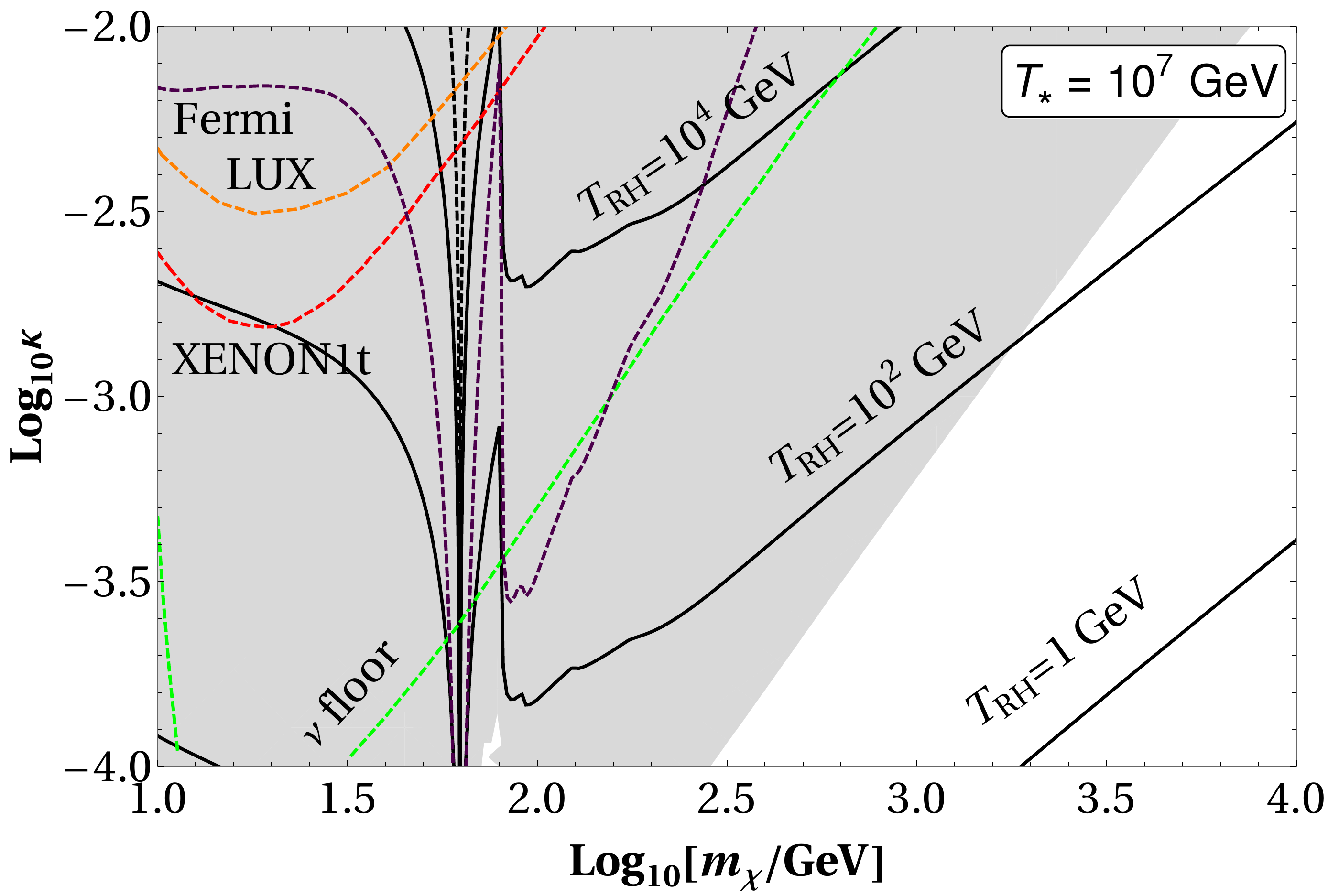}}
	\vspace{-3mm}
	\caption{Experimental constraints on the complex scalar DM annihilating through Higgs portal  from XENON1T \cite{Aprile:2018dbl} (dashed red), LUX \cite{Akerib:2016vxi} (dashed orange), Fermi-LAT \cite{Ackermann:2015zua} (dashed purple), invisible Higgs width \cite{Escudero:2016gzx, Khachatryan:2016whc} (dashed brown). Also shown is the neutrino floor (dashed green). The grey region shows the combined consistency restrictions from Figure~\ref{fig:kContour-ScalarDM}.}
	\label{fig:HiggsExpConScalar}
\end{figure}

Figure \ref{fig:kContour-ScalarDM} indicates that matter dominated freeze-out via the Higgs portal is theoretically viable in modest parameter regions (namely, the white regions), and we now turn to the experimental limits. First we consider direct detection experiments which look for signals from the recoil that a nuclei would undergo in the event that it interacted with DM, and typically provide considerable constraints on DM parameters, see e.g.~\cite{Escudero:2016gzx}. The DM parameters that these experiments exclude for the Higgs model we study may be seen above the red dashed line (XENON1T \cite{Aprile:2018dbl}) and orange dashed line (LUX \cite{Akerib:2016vxi}) lines in Figure \ref{fig:HiggsExpConScalar}. To translate the cross-section bounds of these experiments into Higgs Portal terms, we used the common parameterization for spin-independent (SI) DM-nucleon scattering (see e.g.~\cite{Djouadi:2011aa})
\begin{equation}\label{eq:HPdmNcs}
\sigma_{\chi-N}^{\rm SI}\simeq\frac{\kappa^2}{4\pi m_h^4}\frac{m_N^4f_N^2}{(m_\chi+m_N)^2}~,
\end{equation}
where $m_N\approx1$ GeV is the nucleon mass and a hadronic matrix element parameter $f_N\approx0.326$. 
Note that future direct detection experiments with improved sensitivity will need to confront the difficult task of distinguishing neutrino signals from DM signals in the region termed the ``neutrino floor''. For comparison we indicate the neutrino floor as the green dashed line in Figure \ref{fig:HiggsExpConScalar}. While a nuisance, it is quite consistent for a DM model to have couplings and masses in otherwise allowed parameter space below the neutrino floor curve. 

In addition, there are limits from indirect detection experiments which rely on detecting the products of DM annihilations that may occur. Specifically, the Fermi-LAT  data \cite{Ackermann:2015zua} from observing DM annihilations into the $b\bar{b}$ channel moderately constrains the allowable parameter space for the Higgs portal. The excluded parameters from Fermi-LAT data is shown in Figure \ref{fig:HiggsExpConScalar} as  the dashed purple line. 
Another experimental method for probing DM is through collider searches, if DM interacts with the SM, one would expect that a fraction of the high energy particle collisions would produce DM. By measuring the energy of the products of particle collisions, it may be inferred if there are invisible states which carry away energy undetected. 
 The current branching fraction of the Higgs width that may go to invisible states is constrained by the Large Hadron Collider (LHC) searches to be $\mathcal{B}(h\rightarrow\mathrm{inv})\equiv\Gamma_{\mathrm{inv}}/(\Gamma_{\mathrm{inv}}+\Gamma_{\mathrm{SM}})<0.2$ with 90\% confidence level \cite{Khachatryan:2016whc}. Assuming that the invisible decays are due to DM and using that the Higgs partial decay width to $\chi$ pairs is
\begin{equation}\label{eq:hDecayDMscalar}
\Gamma(h\rightarrow\chi\chi)=\frac{c_{\chi}^{2}\kappa^2v_0^2}{32\pi m_h}\sqrt{1-\frac{4m_\chi^2}{m_h^2}}~,
\end{equation}
implies a constraint on scalar Higgs portal DM \cite{Escudero:2016gzx} which excludes the region above the brown dashed line in Figure \ref{fig:HiggsExpConScalar}.

In Figure \ref{fig:HiggsExpConScalar} we collect both the experimental and theoretical limits, with the theoretical consistency requirements of Figure \ref{fig:kContour-ScalarDM} combined are shown together in grey. While this scenario is quite constrained in the low-mass/high-coupling regime, viable parameter space remains. Notably, radiation dominated freeze-out via the Higgs portal (without an entropy injection) is largely excluded apart from around the region of resonant annihilation as indicated by the black dashed line in Figure \ref{fig:HiggsExpConScalar} (and cf.~\cite{Escudero:2016gzx}). Thus it is interesting to observe that the classic Higgs portals returns as a possibility for providing the correct DM abundance, while avoiding constraints for models with an entropy injection such as  matter dominated DM freeze-out.

It is also worth highlighting that the DM unitarity constraint \cite{Griest:1989wd}, is greatly relaxed in this scenario. Although the limit on plane wave unitarity persists, namely that the maximum cross-section derived from unitarity consideration is $\langle\sigma v\rangle_{\rm max}=4\pi\sqrt{x_{f}}/\sqrt{6}m_\chi^2$, since typically $\Omega_\chi^{\rm relic}h^2\ll \Omega_\chi^{\rm F}h^2$ due to the entropy injection, the freeze-out abundance can be much larger than the standard freeze-out abundance without overclosing the universe.

At this point we should note that the above results are predicated on the assumption that decays of $\phi$ to DM can be neglected. As discussed in Section \ref{s2.3}, while it is not unreasonable to simply assume that $\phi$ has small or absent direct couplings to DM,  induced decays due to Standard Model particles with couplings to both DM and $\phi$ running in loops will lead to an unavoidable non-zero branching ratio $\mathcal{B}_{\rm DM} $ of $\phi$ to DM.

 In the case of the scalar Higgs portal the branching ratio of $\phi$ to DM due to Higgs loops (derived in Appendix \ref{DMdensityPhiDecay}) is given by
\beq
\mathcal{B}_{\rm DM} & \simeq \left[1+\frac{256\pi^{4}}{\kappa^{2}}+\sum_{f}N_c\frac{1024y_{f}^{2}m_{\phi}^{2}\pi^{4}}{\mu_{\phi}^{2}\kappa^{2}}\right]^{-1}\sim \frac{\mu_{\phi}^{2}\kappa^{2}}{3072 \pi^{4} m_{\phi}^{2}}~.
\eeq
where we have  assumed the relevant mass hierarchy $m_\chi, m_h, m_f\ll m_\phi$. Thus parametrically
\beq\label{eq1}
 \mathcal{B}_{\rm DM} & 
\simeq 2\times10^{-9}\left(\frac{\mu_{\phi}}{v_0}\right)^2 \left(\frac{\kappa}{0.1}\right)^2 
\left(\frac{10~{\rm TeV}}{m_{\phi}}\right)^{2}~.
\eeq
We can compare the above to eq.~({\ref{bran}}) which gives the upper bound on  $ \mathcal{B}_{\rm DM}$ such that the loop induced production is subdominant compared to the observed relic density, and thus to the freeze-out abundance in parameter space where the relic density is correctly reproduced.  We observe that for reasonable parameter choices, as those indicated in eq.~(\ref{bran}) \& eq.~({\ref{eq1}}), that $ \mathcal{B}_{\rm DM} $ can be sufficiently small that decays of $\phi$ to DM can be neglected  and that there is a good amount of parameter freedom for which  $\Omega_{\chi, {\rm decay}}h^{2}\ll0.1$.

 \subsection{Fermion dark matter via the Higgs portal}

While the scalar Higgs portal is prized for is minimality, being that it adds only a single state and the interactions proceed via a renormalisable operator  which is generically present in the Lagrangian, there are alternative merits for considering fermion DM. As is well known, the problem with scalars near the electroweak scale is that they typically imply naturalness issues, whereas the relative lightness of fermions is a natural consequence of chiral symmetry~\cite{tHooft:1979rat}. 

Using an  effective field theory (EFT) approach,  fermion DM can be connected to the Standard Model Higgs via the following Lagrangian \cite{Kim:2006af} (see also e.g.~\cite{Freitas:2015hsa,Fedderke:2014wda,Baek:2011aa})
 \beq
 \mathcal{L} = \mathcal{L}_{\text{SM}}+ i\bar{\chi}\gamma^{\mu}\partial_{\mu}\chi - \mu_\chi\bar{\chi}\chi -\frac{1}{\Lambda}H^{\dagger}H\bar{\chi}\chi ~,
 \eeq
where the interaction term of DM with the Higgs involves a dimensionful coupling $\Lambda$. 

\newpage

After the electroweak symmetry breaking, $\chi$ attains a physical mass of $m_{\chi} = \mu_{\chi}+\frac{v_{0}^{2}}{2\Lambda}$, with contributions from both  the bare mass term and the cross-coupling term, leading to
\beq\label{L1}
\mathcal{L} = \mathcal{L}_{\text{SM}}+ i\bar{\chi}\gamma^{\mu}\partial_{\mu}\chi - m_{\chi}\bar{\chi}\chi -\frac{v_0}{\Lambda} h\bar{\chi}\chi - \frac{1}{2\Lambda} h^{2}\bar{\chi}\chi ~.
\eeq
We will mainly consider $m_\chi\gtrsim10$ GeV in which case for $\Lambda\gtrsim 10$ TeV the DM mass is set by the $\mu_\chi$ parameter and is independent of $\Lambda$.
We also define an effective coupling $\kappa = v_{0}/\Lambda$ and thus the decay width of the Higgs boson to the DM can be written as follows
\beq\label{eq:hDecayDMfermion}
\Gamma(h\to\chi\bar{\chi}) = \frac{1}{8\pi }\kappa^{2} m_{h}\left(1-\frac{4m_{\chi}^{2}}{m_{h}^{2}}\right)^{3/2} ~,
\eeq
and the leading annihilation cross-sections in terms of the Mandelstam variable $s$ are
\beq
\sigma_{\chi\bar{\chi}\to f\bar{f}}& =\frac{N_{c}\kappa^2 v_0^2}{16\pi } \frac{m_{f}^{2}\left[s-4m_{\chi}^{2}\right]}{\left(s-m_{h}^{2}\right)^{2}+\left(m_{h}\Gamma_{h}\right)^{2}}\frac{\left[1-\frac{4m_{f}^{2}}{s}\right]^{3/2}}{\sqrt{1-\frac{4m_{\chi}^{2}}{s}}}\\[5pt]
\sigma_{\chi\bar{\chi}\to VV}& =  \frac{\delta_{V}\kappa v_0}{8\pi s } \frac{\sqrt{1-\frac{4m_{V}^{2}}{s}}}{\sqrt{1-\frac{4m_{\chi}^{2}}{s}}}\frac{m_{V}^{4}[s-4m_{\chi}^{2}]}{(s-m_{h}^{2})^{2}+(m_{h}\Gamma_{h})^{2}} \left[2+\frac{(s-2m_{V}^{2})^{2}}{4m_{V}^{4}}\right]~.
\eeq
In the non-relativistic limit, the zeroth order term in the expansion of $\langle\sigma v\rangle$ vanishes and the $p$-wave solution is dominant, implying thermally averaged cross-sections of the form \cite{Kim:2006af} 
\beq\label{eq:fDMannH}
\langle\sigma v\rangle_{\chi\bar{\chi}\to f\bar{f}}& = 
\frac{v^2}{4}\frac{\kappa^2 v_0^2}{4\pi} \frac{r_{f}^{2}(1-4r_{f}^{2})^{3/2}}{\left[(1-r_{h}^{2})^{2}+r_{\Gamma}^{2}\right]}\\[5pt]
\langle\sigma v\rangle_{\chi\bar{\chi}\to VV}& =
\frac{v^2}{4}\frac{\delta_{V}\kappa^2 v_0^2}{8\pi  }\frac{(1+2r_{V}^{2})^{2}}{\left(1-r_{h}^{2}\right)^{2}+r_{\Gamma}^{2}}\sqrt{1-4r_{V}^{2}}~,
\eeq
where $\delta_{V} = 1$ and 1/2 for the $W$ bosons and the $Z$ bosons respectively. 
 
\begin{figure}[t!]
\vspace*{5mm}
\begin{center}
{\bf  Higgs Portal for Fermion Dark Matter}
\end{center}
	\centerline{
	\includegraphics[width=0.6\textwidth]{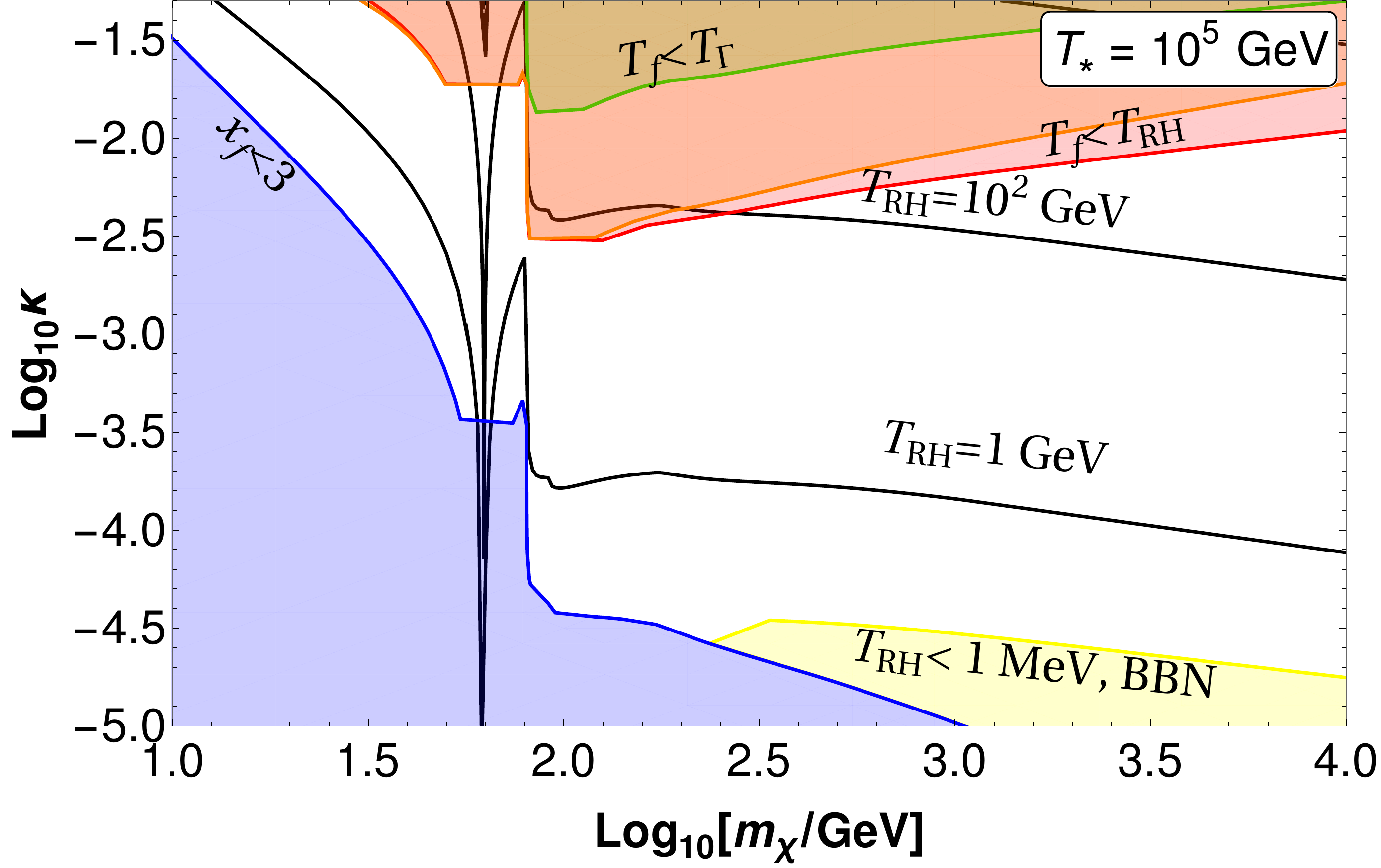}
	\includegraphics[width=0.6\textwidth]{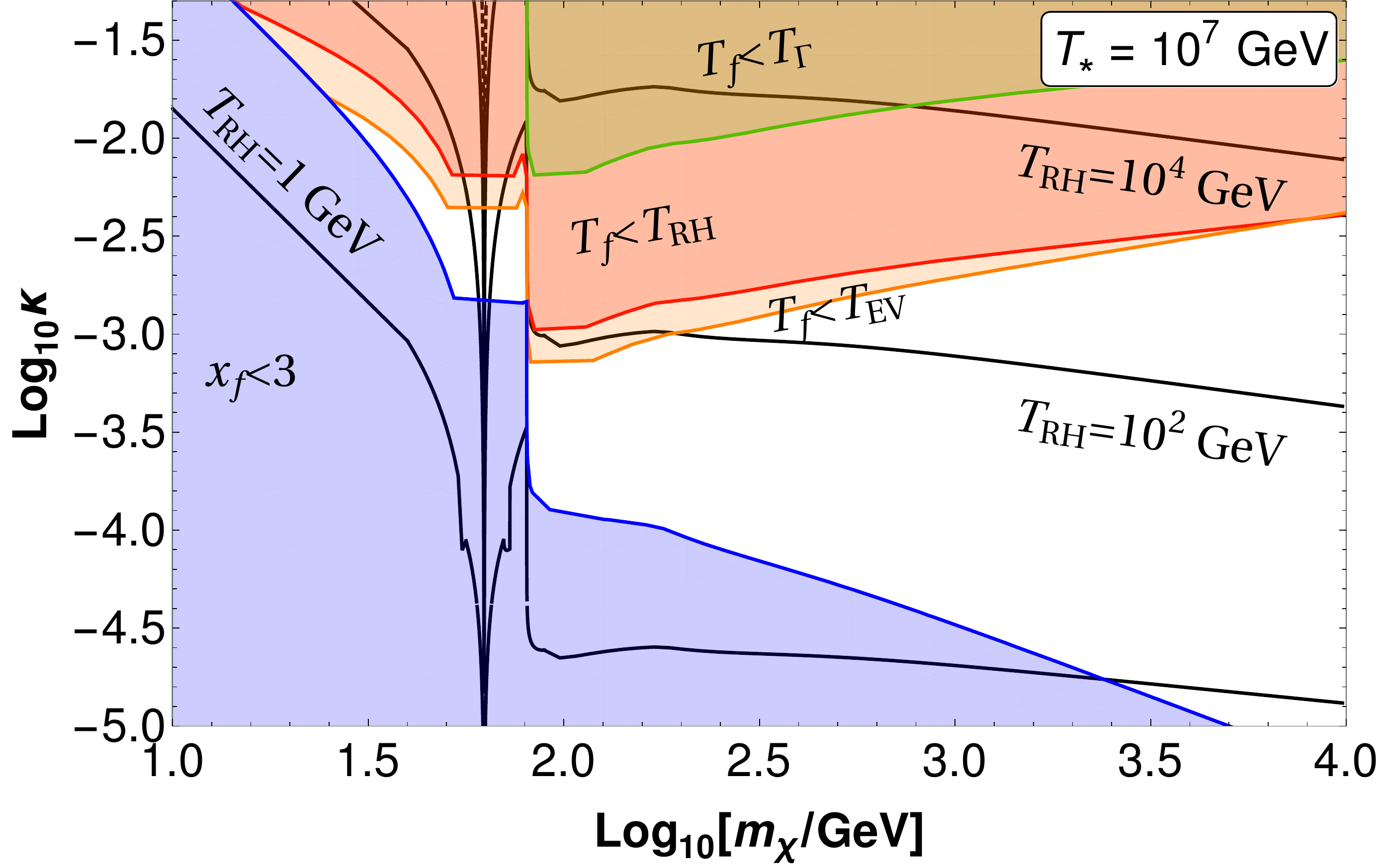}}
\vspace{-1mm}
	\caption{Similar to Figure \ref{fig:HiggsExpConScalar} but for fermion DM annihilating through the EFT Higgs portal for the parameter values $r = 0.99$ and $T_{\star} = 10^{5}$ GeV (left) and $10^{7}$ GeV (right). The solid black curves show contours of the reheat temperature ($T_{\rm RH}$) which give the observed DM relic density. The dashed black line shows the analogous case of radiation dominated DM freeze-out (without an entropy injection). The shaded regions signify parameter values for which the consistency conditions fail: $\phi$ decay before freeze out $T_{f}<T_{\Gamma}$ (green),  decays of $\phi$ cannot be neglected $T_f\ll T_{\rm EV}, T_{\rm RH}$ (red, orange), DM decouples relativistically with $x_f<3$ (blue), and  $T_{{\rm RH}} > 10 $ MeV required for successful BBN  (yellow).  Observe that viable parameter space remains with the observed DM relic density for reheat temperatures less than 1 TeV.}
	\label{fig:kContour-FermionDM}
\end{figure}

 With the above we can calculate the relic abundance due to DM freeze-out for the fermionic Higgs portal, which we show in  Figure \ref{fig:kContour-FermionDM} for two $T_{\star}$ values where we apply the same cosmological consistency conditions  as in Figure \ref{fig:kContour-ScalarDM} (with matching colour scheme).  Unlike in Figure \ref{fig:kContour-ScalarDM} BBN does constrain the parameter space in the plot and the excluded region is shaded in yellow.
There is a considerable amount of viable parameter space with regards to the consistency conditions and thus we next look at the experimental constraints.  

\begin{figure}[t!]
	\centerline{\includegraphics[width=0.6\textwidth]{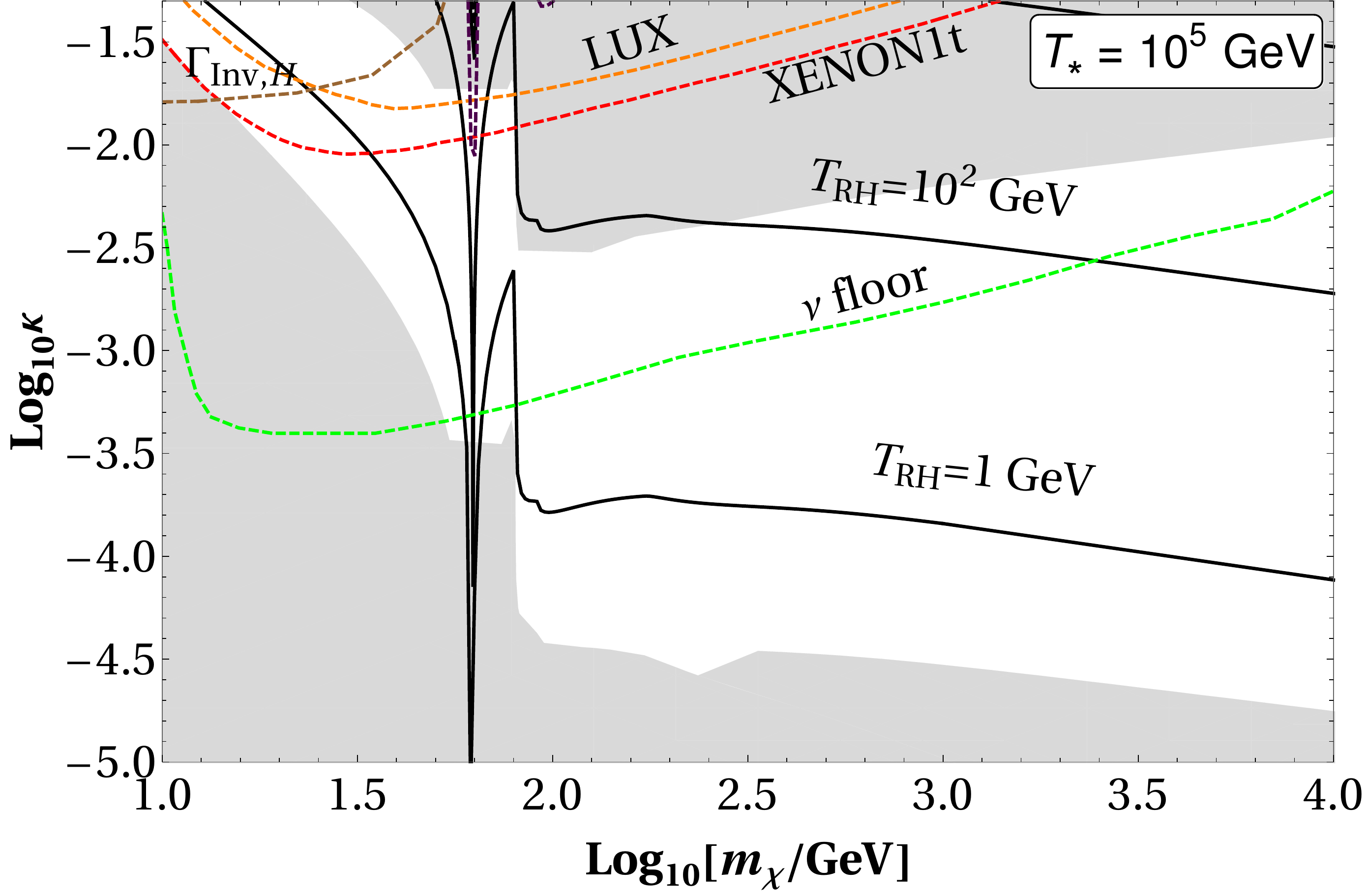}
	\includegraphics[width=0.6\textwidth]{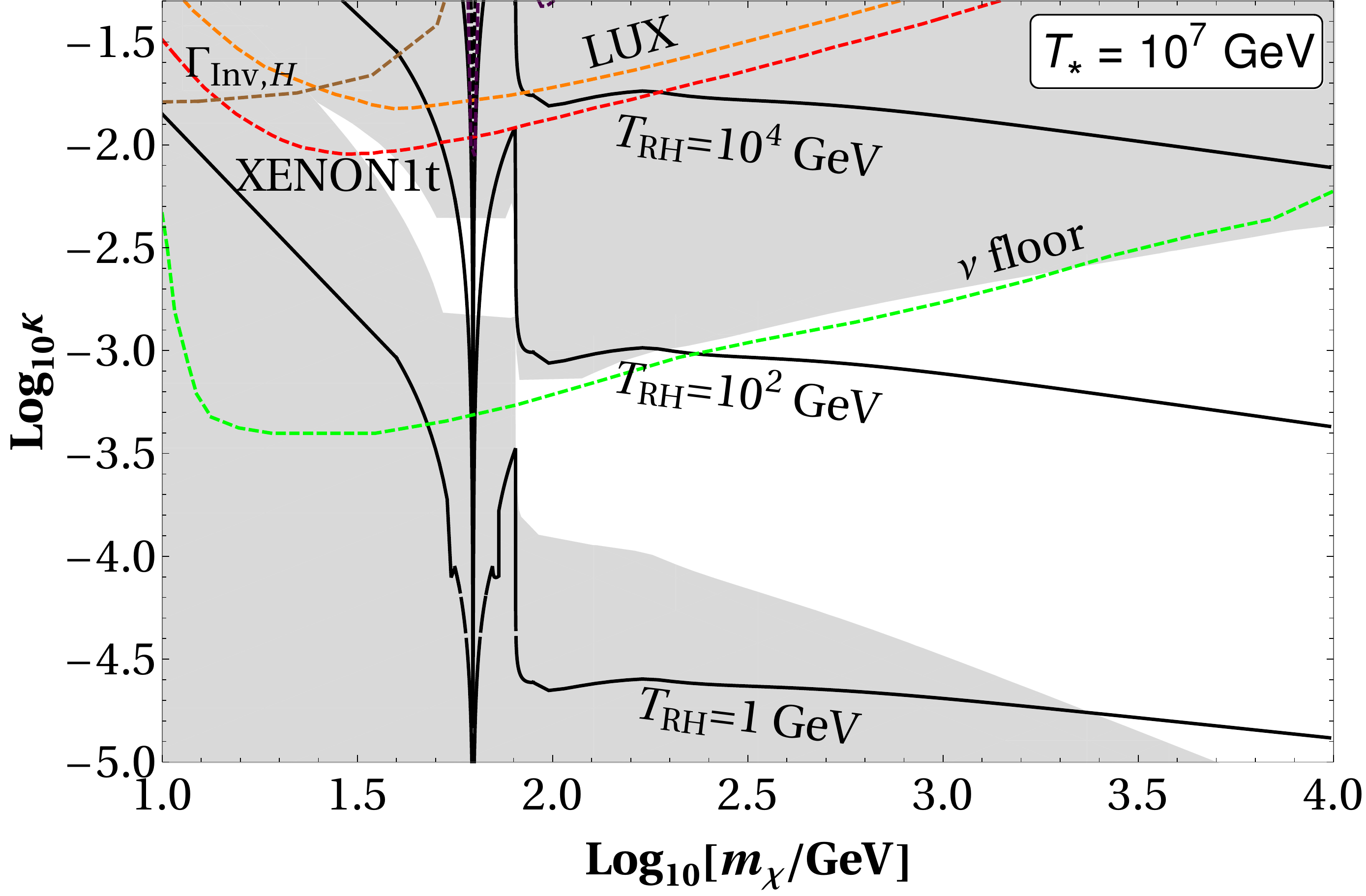}}
		\vspace{-1mm}
\caption{Experimental constraints on the fermion DM annihilating through Higgs portal  from XENON1T \cite{Aprile:2018dbl} (dashed red), LUX \cite{Akerib:2016vxi} (dashed orange), Fermi-LAT \cite{Ackermann:2015zua} (dashed purple), invisible Higgs width \cite{Escudero:2016gzx, Khachatryan:2016whc} (dashed brown). Also shown is the neutrino floor (dashed green line). The grey region shows the combined consistency restrictions from Figure~\ref{fig:kContour-FermionDM}.}
	\label{fig:HiggsExpConFermion}
\end{figure}

While the experiments which constrain this model  are similar to the scalar Higgs portal, the exclusion curves differ considerably. Fermion DM coupling with the Higgs boson via eq.~(\ref{L1}) leads to a spin independent direct detection cross-section and here we follow \cite{Baum:2017enm} taking
  \beq\label{fermionHPDDcs}
  \sigma^{\rm SI}_{p,h}\sim 5\times 10^{-11}\times {\rm pb} \left(\frac{\kappa}{0.01}\right)^{2}~.
  \eeq
We present the parameter space of the fermion Higgs portal constrained by the various experimental limits in Figure \ref{fig:HiggsExpConFermion}. We highlight that a considerable amount of parameter space is viable and not currently excluded by searches, although for larger values of $T_\star$ much of it falls beneath the neutrino floor (green dashed curve) making detection challenging.

\vspace*{8mm}

\clearpage
\newpage

Despite the theoretical and experimental constraints on the parameters, a considerable range of parameter space remains allowed for   Higgs portal fermion DM in matter dominated freeze-out.  Notably, for $T_\star<10^7$ GeV there is a wide range of acceptable DM masses with $1~{\rm GeV}\lesssim T_{\rm RH}\lesssim100$ GeV which give the correct relic density with a cross-section which is consistent with matter dominated freeze-out. Moreover, many acceptable  parameter points  are above the neutrino floor and thus potentially discoverable by future direct detection experiments. An interesting region in Figure \ref{fig:HiggsExpConFermion} is $m_{\chi}>100$ GeV where the dominating annihilation channel of the DM is to the $W$ bosons, and in this case DM freeze-out occurring during matter domination decouples at
\beq
x_{f} \simeq \ln\left(\sqrt{\frac{45}{7\pi^{8}}}\frac{gM_{{\rm Pl}}\kappa^{2}}{v_{0}^{2}g_*^{1/2}T_{\star}^{1/2}x_{f}^{3/2}}(1-r)^{-1/2}\frac{m_{\chi}^{3/2}\left(1+\frac{m_{W}^{2}}{2m_{\chi}^{2}}\right)^{2}\sqrt{1-\frac{m_{W}^{2}}{m_{\chi}^{2}}}}{\left(1-\frac{m_{W}^{2}}{4m_{\chi}^{2}}\right)^{2}+\left(\frac{m_{h}\Gamma_{h}}{4m_{\chi}^{2}}\right)^{2}}\right)~,
\eeq
It follows that for $m_{\chi}>100$ GeV  DM relic density  can be approximated succinctly   as
\beq
\Omega_{{\rm DM}}^{{\rm Relic}}h^{2} &\simeq
5\times 10^{10}~{\rm GeV}^{-1}  \times\zeta \frac{(1-r)^{1/2}g_*^{-1/2}}{\delta_{V}\kappa^4 v_0^2M_{{\rm Pl}}}\frac{x_{f}^{5/2}}{x_{\star}}\frac{\left[\left(1-\frac{m_{h}^{2}}{4m_{\chi}^{2}}\right)^{2}+\left(\frac{m_{h}\Gamma_{h}}{4m_{\chi}^{2}}\right)^{2}\right]}{\left(1+\frac{m_{W}^{2}}{2m_{\chi}^{2}}\right)^{2}\sqrt{1-\frac{m_{W}^{2}}{m_{\chi}^{2}}}}~.
\eeq

Finally, we discuss the limits on the parameter space which come from requiring that the loop induced decays of $\phi$ to DM lead to a subdominant contribution to the relic density compared to the population from matter dominated freeze-out.
Taking the relevant mass ordering $m_\chi, m_h \ll m_\phi$ the branching ratio to DM in this case is (see Appendix \ref{DMdensityPhiDecay})
\beq
\mathcal{B}_{\rm DM}
&\simeq \frac{m_{\phi}^{2}}{64\Lambda^{2}\pi^{4}}\sim 10^{-8}\left(\frac{m_\phi}{10^4~{\rm GeV}}\right)^2\left(\frac{10^6~{\rm GeV}}{\Lambda}\right)^2~.
\eeq
Small branching to DM requires a sizeable separation $m_\phi\ll \Lambda$ but this then also implies a suppression in the effective coupling to the Higgs $\kappa= v_0/\Lambda$ since we assume that $m_\phi>m_h$.  For the reference values above, the corresponding effective couplings are of order $\kappa\sim3\times10^{-4}$.  On the other hand, matter dominated freeze-out via the fermion Higgs portal with $m_\chi>$100 GeV and a reheat temperature above BBN requires a coupling $\kappa\gtrsim3\times 10^{-5}$ (cf.~Figure \ref{fig:kContour-FermionDM}) which implies a relatively low EFT cut off $\Lambda$, and thus to keep $\mathcal{B}_{\rm DM}$ negligible $m_\phi$ must also be light. However, inspection to eq.~({\ref{bran}}) reveals that reducing $m_\phi$ increases the production to DM, and thus one needs a smaller branching ratio. This compresses the viable model space but for sub-PeV $\phi$ it can be arranged for $\mathcal{B}_{\rm DM}$ to be sufficiently small, while simultaneously allowing appropriate couplings for matter dominated freeze-out and $T_{\rm RH}\gtrsim10$ MeV.


\section{Matter Dominated Freeze-out via $Z$-mediation}
\label{s4}

One of the principle reasons that Higgs portal models are deemed compelling extensions of the Standard Model which include DM is because of their extreme minimality. There is another class of models which shares this elegant feature, namely $Z$ mediated DM, in which the DM carries electroweak quantum numbers and interacts via the Standard Model $Z$ boson. While it is an elegant manner of connecting DM to the known particles, it suffers the same set back to the Higgs portal in that $Z$ mediated freeze-out is largely excluded by experimental searches in the case of radiation domination. In this section we reconsider this classic model in the context of matter dominated freeze-out and show that viable parameter ranges persist. 

\newpage
\subsection{Vector-like fermion dark matter}

We shall consider here the case of fermion DM which carries a Standard Model electroweak charge and thus interacts with $Z$ boson via the following Lagrangian terms
\begin{align}
\label{fDMZportal}
& \mathcal{L}\supset\frac{g}{4 \cos\theta_W}\left( \overline{\chi}\gamma^{\mu} \left(V_{\chi}-A_{\chi} \gamma^{5}\right) \chi Z_{\mu}\right)~,
\end{align}
\noindent
where $g\simeq 0.65$ is the electroweak coupling, $ V_{\chi}$ and $ A_{\chi}$  are the DM vector and axial couplings to the $Z$ boson and $\theta_{W}$ is the weak mixing angle. 
It follows that the decay width of the $Z$ boson to the DM is given by \cite{Arcadi:2014lta}
\beq\label{eq:ZdecayfDM}
\Gamma(Z\to \chi\bar{\chi})= \frac{g^2}{192 \pi \cos^2 \theta_W} m_Z 
\sqrt{1 - \frac{4 m_\chi^2}{m_Z^2}} \left[|V_\chi|^2  \left( 1 + \frac{2 m_\chi^2}{m_Z^2} \right) + |A_\chi|^2 \left( 1 - \frac{4 m_\chi^2}{m_Z^2} \right) \right]~.
\eeq
The thermally averaged annihilations cross-sections of the fermion DM via the $Z$ portal are given in Appendix \ref{apa}. The couplings  $V_{\chi}$ and $A_{\chi}$ depend on the model, as we discuss shortly.

For $V_\chi\neq0$ SI scattering constrains this scenario and the relevant cross-sections are \cite{Arcadi:2014lta}
\beq\label{sicr}
&\sigma^{p}_{\rm SI} = \frac{g^{4}|V_{\chi}|^{2}\mu_{\chi p}^{2}}{4\pi m_{Z}^{4}\cos^{2}\theta_{W}}\frac{\sum_{A}\eta_{A}A\left[V_{u}\left(1+Z/A\right)+V_{d}\left(2-Z/A\right)\right]^{2}}{\sum_{A}\eta_{A}A^{2}}~,\\
&\sigma^{n}_{\rm SI} = \sigma^{p}_{\rm SI}\frac{\mu_{\chi n}^{2}}{\mu_{\chi p}^{2}}\frac{(V_{u}+2V_{d})^{2}}{(2V_{u}+V_{d})^{2}}~,
\eeq
where $Z$ and $A$ are respectively the atomic and nucleon numbers and $\mu_{\chi i} = \frac{m_{\chi}m_{i}}{m_{\chi}+m_{i}}$ is the reduced mass. The quantity $\eta_A$ denotes the relative abundance of the target material. Specifically, for Xenon1T the two main Xenon isotopes are $^{129}{\rm Xe}$ (Spin 1/2) and $^{131}{\rm Xe}$ (Spin-3/2) with relative abundances $\eta_{129}=0.264$ and $\eta_{131}=0.212$ respectively \cite{Aprile:2019dbj}.  In the above $V_f$ and $A_f$ are the ordinary vector and axial couplings of the $Z$ boson to the Standard Model fermions $f$  carrying charge $q_{f}$, with third component of isospin $T^{3}_{f}$ defined as
\beq
V_{f} = 2\left(-2q_{f}\sin^{2}\theta_{W}+T^{3}_{f}\right)~,
\qquad
A_{f} = 2T^{3}_{f}~.
\eeq

\begin{figure}[t!]
\vspace*{10mm}
\begin{center}
{\bf  $Z$ Portal for Vector-like Fermion Dark Matter }
\end{center}
\vspace*{6mm}
	\centerline{
	\includegraphics[width=0.6\textwidth]{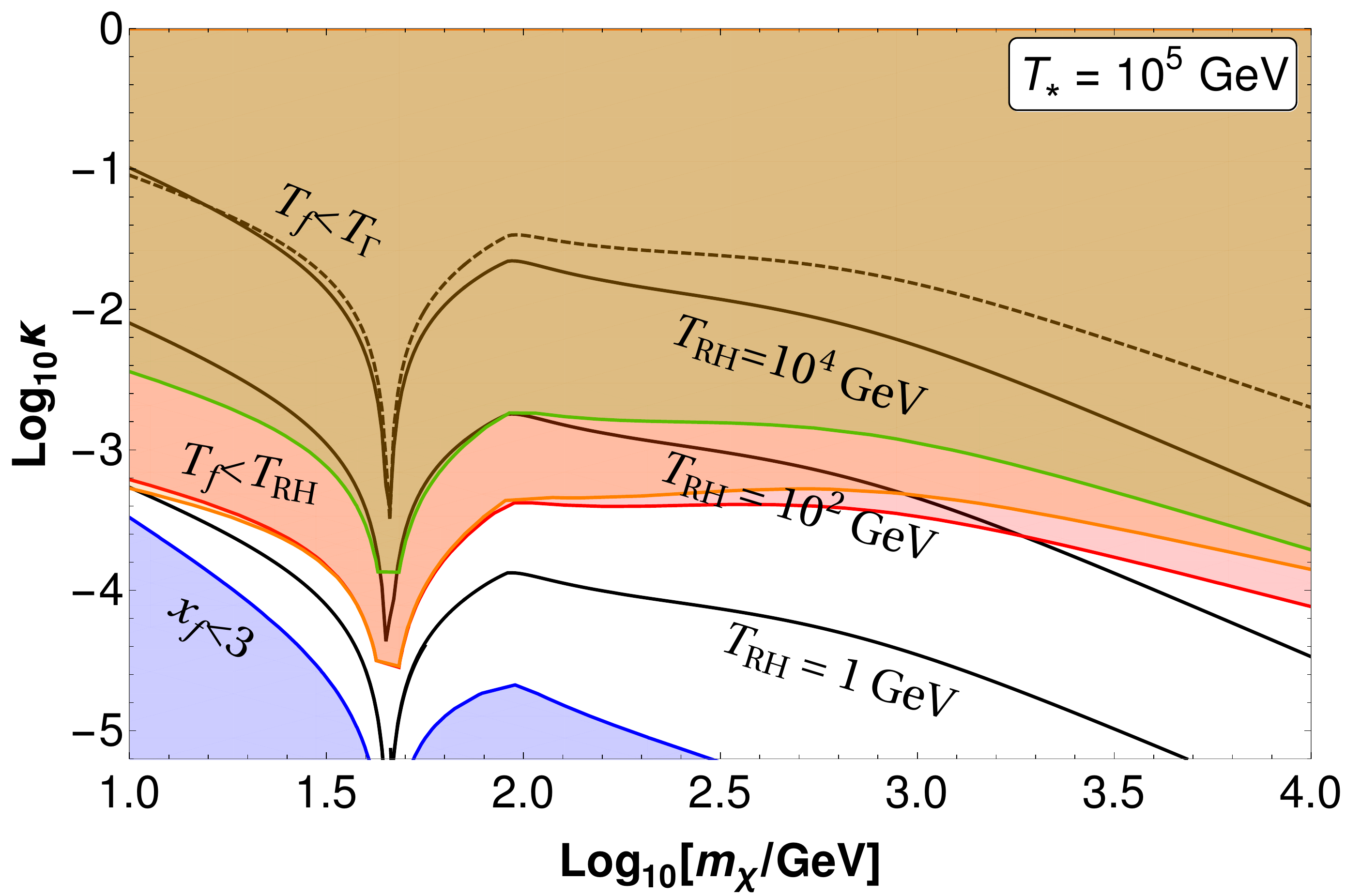}
	\includegraphics[width=0.6\textwidth]{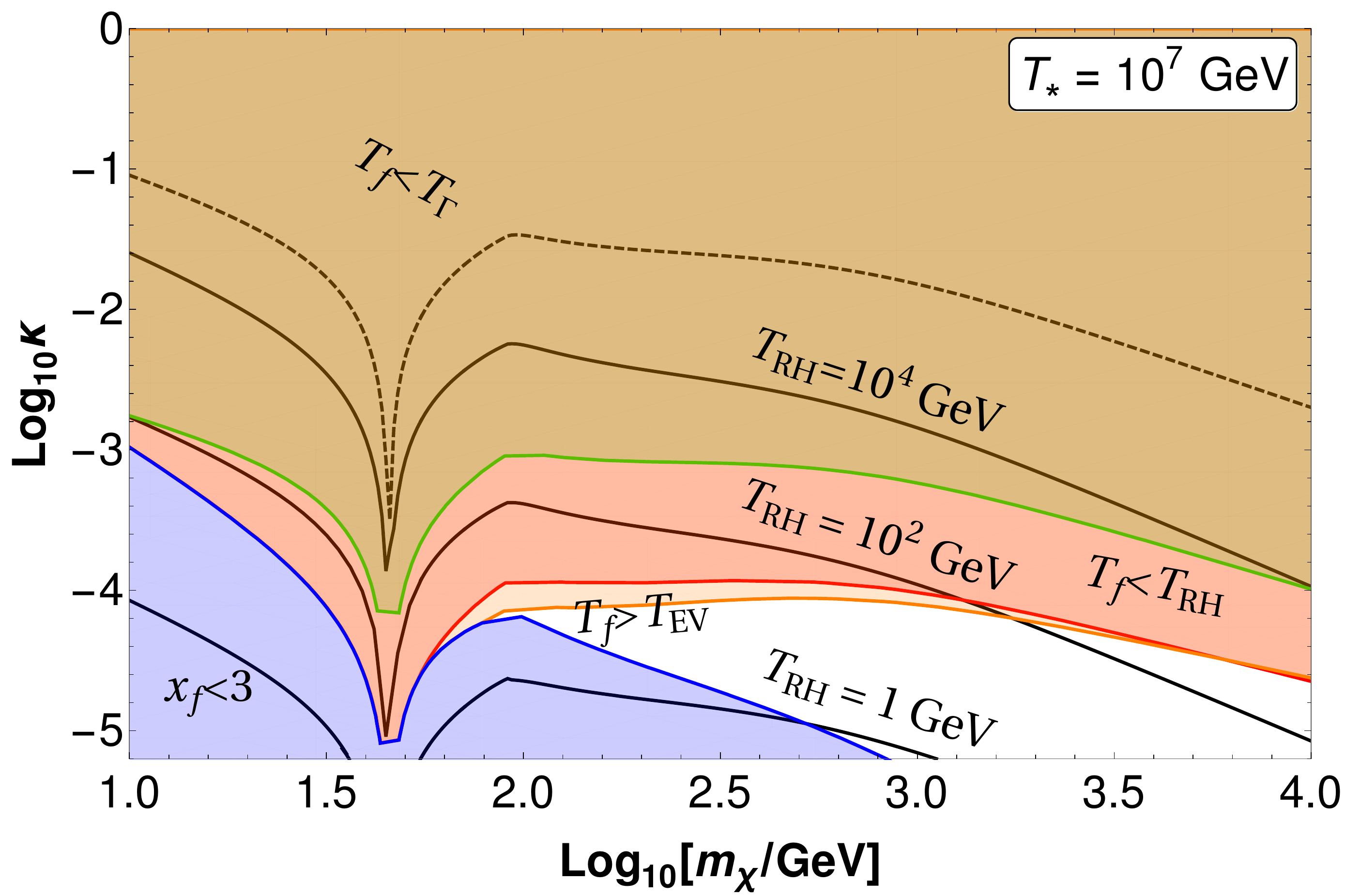}}
		\vspace{2mm}
	\caption{Similar to Figure \ref{fig:HiggsExpConScalar} but for vector-like fermion DM annihilating via the $Z$ portal. The solid black curves show contours of the reheat temperature ($T_{\rm RH}$) in the matter dominated DM freeze-out scenario for a complex scalar DM annihilating through Higgs portal which give the observed DM relic density taking the parameter values $r = 0.99$ and $T_{\star} = 10^{5}$ GeV (left) and $10^{7}$ GeV (right). The dashed black line shows the analogous case of radiation dominated DM freeze-out (without an entropy injection).}
	\label{fig:kContour-FermionDMZportal}
\end{figure}

\begin{figure}[t!]
	\centerline{\includegraphics[width=0.6\textwidth]{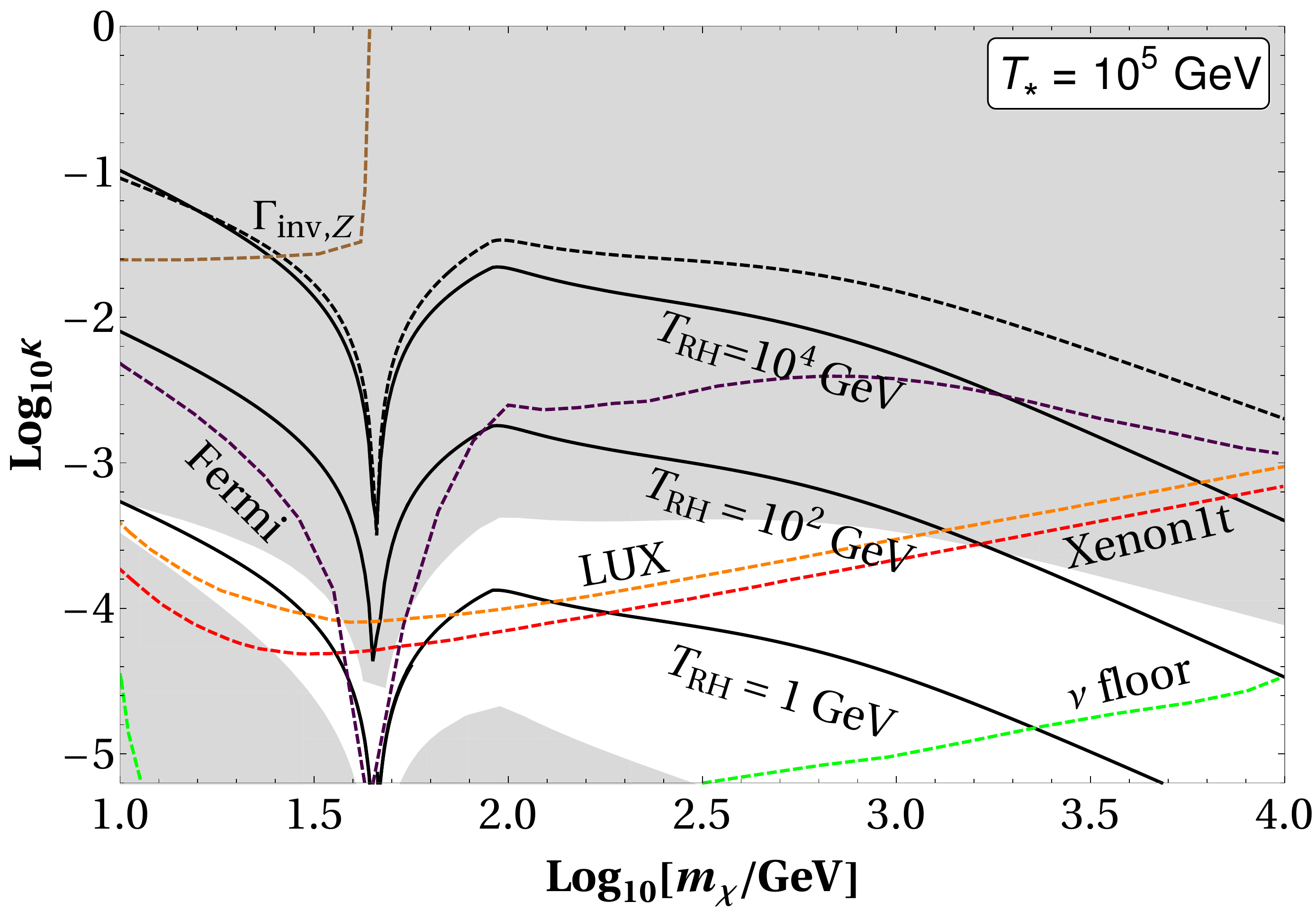}~~
	\includegraphics[width=0.6\textwidth]{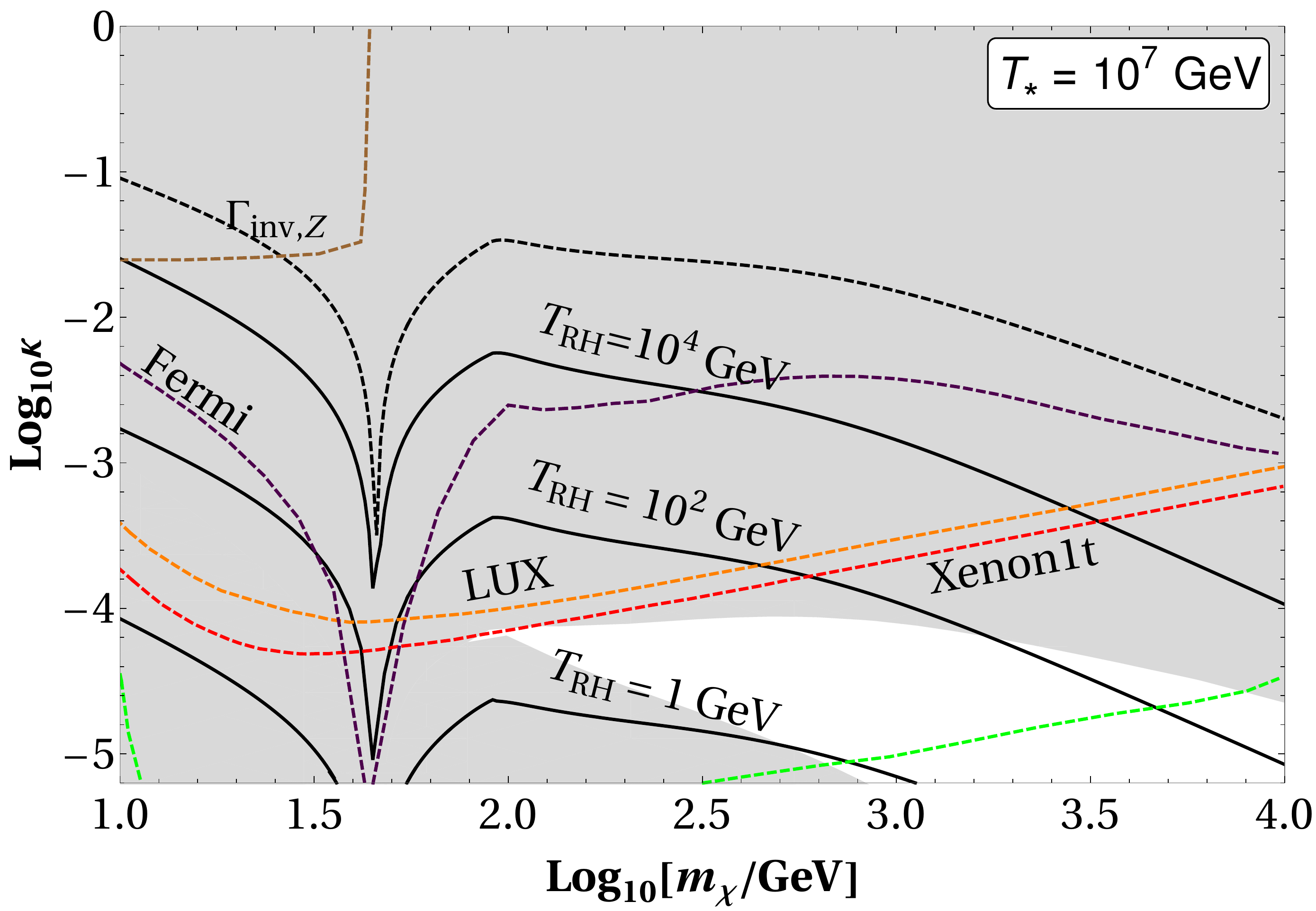}}
		\vspace{2mm}
\caption{Experimental constraints on vector-like fermion DM annihilating through the $Z$ portal from XENON1T \cite{Aprile:2018dbl} (dashed red), LUX \cite{Akerib:2016vxi} (dashed orange), Fermi-LAT \cite{Ackermann:2015zua} (dashed purple), invisible $Z$ width \cite{Escudero:2016gzx, Khachatryan:2016whc} (dashed brown). Also shown is the neutrino floor (dashed green line). The grey region shows the combined consistency restrictions from Figure~\ref{fig:kContour-FermionDMZportal}.
		\label{fig:ZbosonExpConFermion}}
\end{figure}

The most minimal case is that the DM fermions are vector-like (i.e.~non-chiral), which implies that the $\chi_L$ and $\chi_R$ have the same quantum numbers, resulting in $V_\chi=A_\chi$ and we relabel this coupling $\kappa$. Notably, these states cancel trivially in their additional contributions to the gauge anomalies of the Standard Model group.  Similar to the previous plots, in Figures \ref{fig:kContour-FermionDMZportal} \& \ref{fig:ZbosonExpConFermion}  we show the DM relic density in the $\kappa$-$m_{\chi}$ plane and constrain it with the consistency conditions and  experimental constraints. 

\subsection{Axially coupled fermion dark matter}

The case $V_\chi\neq0$ generically leads to strong SI limits, and the most minimal scenario involving fermion DM coupling via the $Z$ boson with $V_\chi=A_\chi$ for radiation dominated freeze-out. This has motivated previous groups (e.g.~\cite{Arcadi:2014lta,Escudero:2016gzx}) to consider $V_\chi=0$ in which case the leading SI cross-section of eq.~(\ref{sicr}) is zero. We note that the axial coupling implies that the DM must be chiral which naively leads to gauge anomalies, but these can be cancelled via the inclusion of additional states \cite{Ismail:2016tod,Ellis:2017tkh}. Such constructions tend to be highly non-minimal however and thus we view the vector-like case as much more compelling.

\newpage

\clearpage
\newpage

\clearpage
\newpage
\begin{figure}[b!]
\vspace*{10mm}
\begin{center}
{\bf  $Z$ Portal for Axially Coupled Fermion Dark Matter }
\end{center}
\vspace*{6mm}
	\centerline{
	\includegraphics[width=0.6\textwidth]{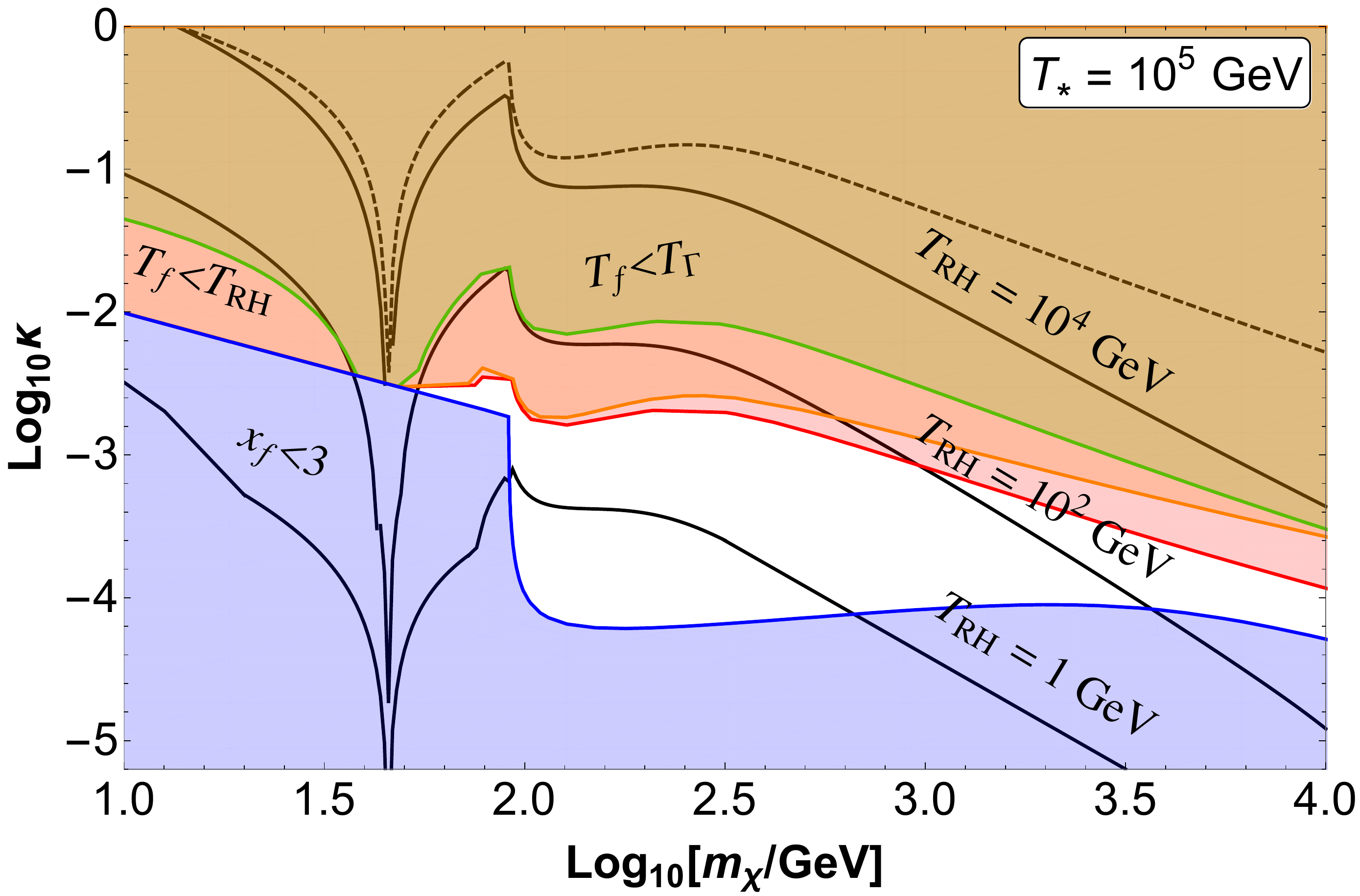}
	\includegraphics[width=0.6\textwidth]{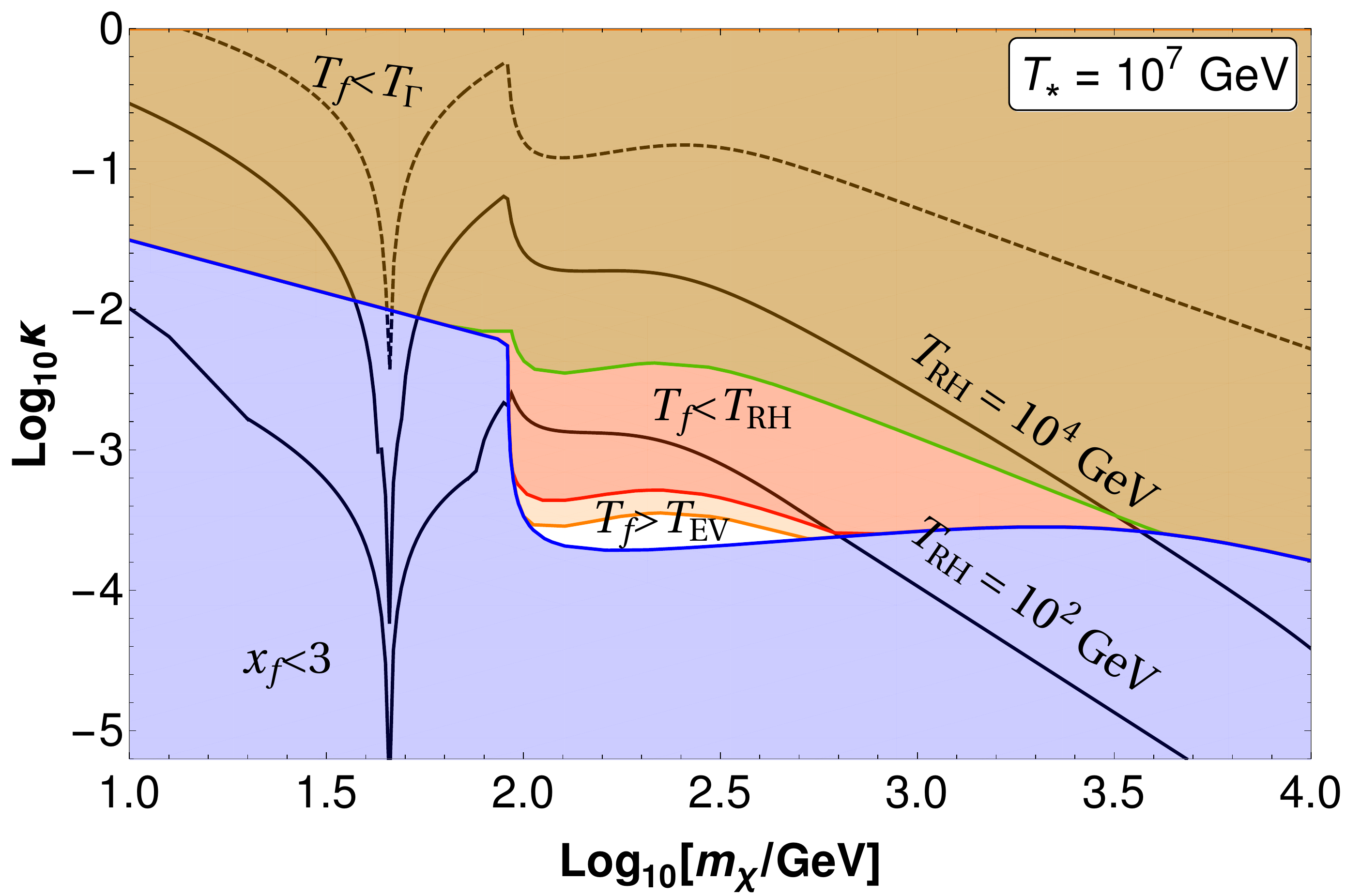}}
		\vspace{2mm}
	\caption{Similar to Figure \ref{fig:HiggsExpConScalar} but for axially-coupled fermion DM  annihilating via the $Z$ portal. The solid black curves show contours of the reheat temperature ($T_{\rm RH}$) in the matter dominated DM freeze-out scenario for a complex scalar DM annihilating through Higgs portal which give the observed DM relic density taking the parameter values $r = 0.99$ and $T_{\star} = 10^{5}$ GeV (left) and $10^{7}$ GeV (right). The dashed black line shows the analogous case of radiation dominated DM freeze-out (without an entropy injection).}
	\label{fig:axial1}
\end{figure}

\begin{figure}[b!]
	\centerline{\includegraphics[width=0.6\textwidth]{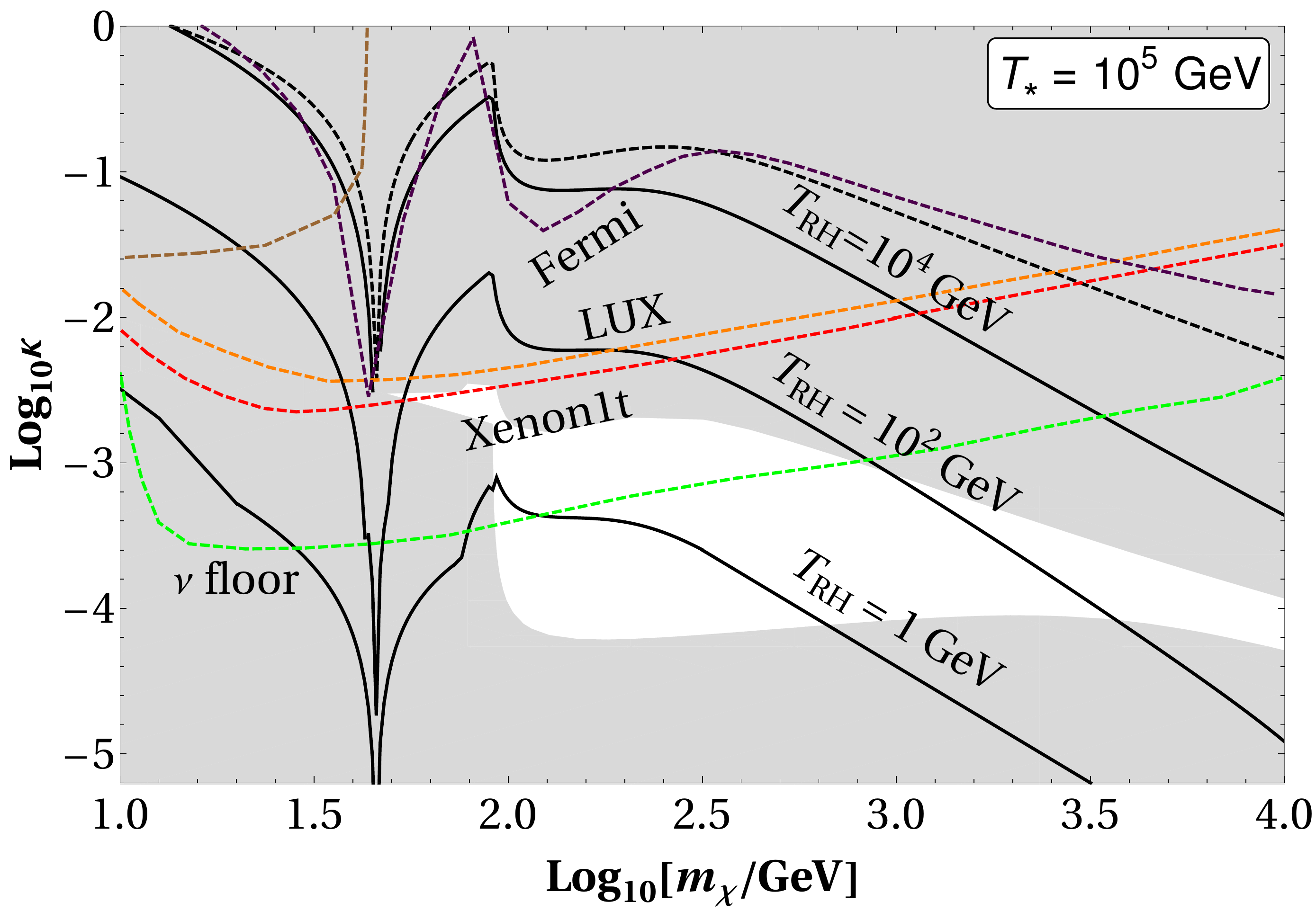}
	\includegraphics[width=0.6\textwidth]{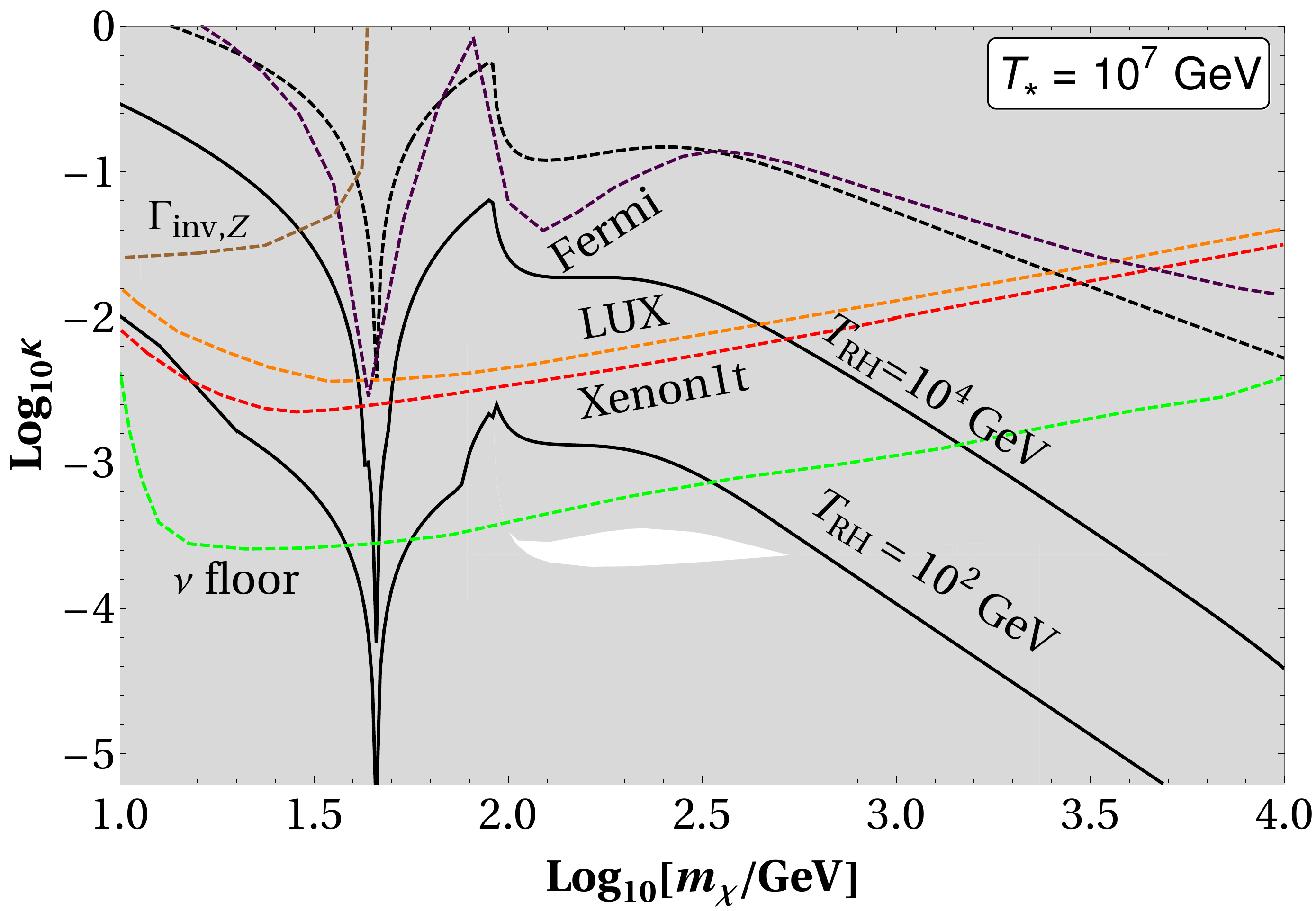}}
		\vspace{2mm}
		\caption{Experimental constraints on axially-coupled fermion DM annihilating through the $Z$ portal from XENON1T \cite{Aprile:2018dbl} (dashed red), LUX \cite{Akerib:2016vxi} (dashed orange), Fermi-LAT \cite{Ackermann:2015zua} (dashed purple), invisible $Z$ width \cite{Escudero:2016gzx, Khachatryan:2016whc} (dashed brown). Also shown is the neutrino floor (dashed green line). The grey region shows the combined consistency restrictions from Figure~\ref{fig:axial1}. \label{fig:axial2}}
\end{figure}
\clearpage
\newpage

With $V_\chi=0$ and $A_\chi\neq0$  the (much weaker)  spin dependent  (SD) limits provide the leading constraints. The  scattering cross-sections  relevant for SD direct detection are
\beq
&\sigma^{p}_{\rm SD} = \frac{3g^{4}|A_{\chi}|^{2}\mu_{\chi p}^{2}}{4\pi m_{Z}^{4}\cos^{2}\theta_{W}}\frac{\sum_{A}\eta_{A}\left[A_{u}\left(\Delta^{p}_{u}S^{A}_{p}+\Delta^{p}_{d}S_{n}^{A}\right)+A_{d}\left(S_{p}^{A}\left(\Delta^{p}_{d}+\Delta^{p}_{s}\right)+S_{n}^{A}\left(\Delta^{p}_{u}+\Delta^{p}_{s}\right)\right)\right]^{2}}{\sum_{A}\eta_{A}(S^{A}_{p}+S^{A}_{n})^{2}}~,\\[8pt]
&\sigma^{n}_{\rm SD} = \sigma^{p}_{\rm SD}\frac{\mu_{\chi n}^{2}}{\mu_{\chi p}^{2}}\frac{\left(A_{u}\Delta_{u}^{n}+A_{d}\left(\Delta_{d}^{n}+\Delta_{s}^{n}\right)\right)^{2}}{\left(A_{u}\Delta_{u}^{p}+A_{d}\left(\Delta_{d}^{p}+\Delta_{s}^{p}\right)\right)^{2}}~,
\eeq
where  the nucleon spin-content parameters are \cite{Ellis:2008hf}:
$\Delta^{p}_{u} = \Delta^{n}_{d}, ~~\Delta^{p}_{d} = \Delta^{n}_{u}, ~~\Delta^{p}_{s} = \Delta^{n}_{s},$ \linebreak
$\Delta^{p}_{u} = 0.84, ~~ \Delta^{p}_{d} = -0.43, ~~ \Delta^{p}_{s} = -0.09.$
The dominant Xenon isotopes that contribute to the SD interactions imply different expectation values of proton and neutron spin operators  \cite{Klos:2013rwa} with
$^{129}S_{n} = 0.329,~ ^{129}S_{p} = 0.01, ~  ^{131}S_{n}=-0.272,$ and $^{131}S_{p}=-0.009.$

In Figures \ref{fig:axial1} \& \ref{fig:axial2} we show the consistency and experimental constraints for the axially coupled case where we rename the single coupling $\kappa=A_\chi$.  While the direct  and indirect detection limits are relaxed, the cosmological consistency conditions are substantially more constraining. Note that for $T_\star\sim 10^7$ GeV almost no consistent parameter space remains, however for smaller values of $T_\star$ (and this also  depends on the initial condition on $r$) then more viable parameter space is available and for $T_\star\sim 10^5$ GeV DM between 100 GeV and 10 TeV can be accommodated.  Finally, we note that since $\phi$ is assumed to be a gauge singlet, it does not couple directly to $Z$. Therefore the loop induced production of DM from $\phi$ decays is at two loops and thus are significantly less restrictive than the scalar Higgs portal which are already largely unconstraining, as discussed previously.

\section{Conclusions}
\label{s5}

Classic scenarios in which DM couples to the Standard Model in a minimalistic fashion involving the Higgs or Standard Model $Z$ boson offer an extremely compelling  picture of nature. However, under the assumption that DM decouples during radiation domination these elegant DM scenarios are essentially excluded by experimental searches. Here we highlighted that these classic models can be revived if the DM decouples during an early period of matter domination which subsequently transitions to radiation domination prior to BBN. Specifically, in the context of the scalar Higgs portal, the fermion (EFT) Higgs portal, and the $Z$ portal for fermion DM, we outlined the consistency conditions for matter dominated freeze-out and explored the leading experimental limits which constrain these scenarios. 

The main results of this paper are encapsulated  in Figures \ref{fig:HiggsExpConScalar},  \ref{fig:HiggsExpConFermion}, and  \ref{fig:ZbosonExpConFermion} which show slices of the viable parameter space for the scalar and fermion Higgs portals, and the $Z$ portal. The white space in these plots indicates the parameter space consistent with the assumptions of matter dominated freeze-out and the coloured dashed lines indicate the upper limits on the (effective) coupling. Thus the viable parameter space is the white regions below all of the coloured lines (except the neutrino floor curve).  In each of the DM portals explored here, there exists viable parameter space below the experimental exclusions and above the neutrino floor. Since the next generation of detectors will be able to reach the neutrino floor, this indicates that these scenarios are potentially testable in the near future. Given that the modifications needed to the standard cosmology are minimal in order to revive these classic models, and that these scenarios can be probed in the next decade, this presents an exciting prospect for discovery.

It would be interesting to re-examine other motivated models which face strong experimental limits within the context of matter dominated DM freeze out. In particular, there are several interesting variants of the Higgs portals,  for instance involving supersymmetry \cite{MarchRussell:2008yu}, two Higgs doublets \cite{Bell:2017rgi}, or vectorial DM \cite{Lebedev:2011iq}. 
Another related scenario is the case of DM coupling via a non-Standard Model abelian gauge boson $Z'$, see e.g.~\cite{Alves:2013tqa,Blanco:2019hah,Arcadi:2013qia}. It has been proposed that such a $Z'$ could couple to DM and some Standard Model states, while the ordinary $Z$ has only Standard Model interactions.  In particular, one might consider a sequential $Z'$, i.e.~a heavier version of the Standard Model $Z$ with matching couplings strengths. Notably, if DM decouples during an early period of matter domination, then search bounds \cite{Blanco:2019hah} will be ameliorated similar to the DM portals explored here.


\subsection*{Acknowledgements} 
We thank Nicolas Bernal, Chris Kolda, and Stephen West for helpful interactions. 
JU and PC are grateful to the Simons Center for Geometry and Physics (Program: Geometry \& Physics of Hitchin Systems)  for their hospitality and support.  JU acknowledges support from the National Science Foundation grant  DMS-1440140 while in residence at the Mathematics Sciences Research Institute, Berkeley during Fall 2019.

\appendix

\section{Thermally averaged annihilation cross-section fermion dark matter  via~$Z$ }
\label{apa}

In this appendix we provide for reference the thermally averaged cross-sections for the DM annihilating to the Standard Model as used in Section \ref{s4}. We follow the treatment given in \cite{Arcadi:2014lta}. The thermally averaged annihilation cross-section of DM to Standard Model fermions can be expressed as
\begin{align}\label{eq:fDMannZ1}
& \langle \sigma v \rangle_{\chi\bar{\chi}\to f\bar{f}} = \sum_{m_{f}<m_{\chi}}I_{0}^{f}\left[I_{s}^{f}+v^{2}I_{p}^{f}\right]~.
\end{align}
where $v= \sqrt{3T/m_{\chi}}$ is the velocity and this is parameterised in terms of an overall factor $I_{0}^{f}$, an $s$-wave factor $I_{s}^{f} $ and a $p$-wave factor $I_{p}^{f}$. The overall factor is given by
\beq
I_{0}^{f} = \frac{g^{4}\kappa^{2}}{384\pi\cos^{2}\theta_{W}}N_{c}^{f}\frac{1}{m_{\chi}m_{Z}^{4}\sqrt{m_{\chi}^{2}-m_{f}^{2}}\left(m_{Z}^{2}-4m_{\chi}^{2}\right)^{3}} ~.
\eeq
In terms of  $\hat{A}_{\chi}=A_{\chi}|_{\kappa =1}$ and $\hat{V_{\chi}}=V_{\chi}|_{\kappa =1} $ the $s$-wave factor is as follows
\beq
I_{s}^{f}  = 12 \left(m_{\chi}^{2}-m_{f}^{2}\right)\left(m_{Z}^{2}-4m_{\chi}^{2}\right) \left(|\hat{A}_{\chi}|^{2}J^{f}_{sA}+ |\hat{V}_\chi|^2 J^{f}_{sV}\right) ~,
\eeq
where
\beq
J^{f}_{sV}&=m_{Z}^{4} \left(2 |A_f|^2 \left(m_\chi^2-m_f^2\right)+|V_f|^2 \left(m_f^2+2 m_\chi^2\right)\right)
\eeq
and
\beq
J^{f}_{sA}&=|A_f|^2  m_f^2 \left(m_{Z}^2-4 m_\chi^2\right)^2~.
\eeq
Finally, the $p$-wave factor is of the form
\beq
I_{p}^{f} =- \left[|\hat{V_{\chi}}|^{2}\left(|A_{f}|^{2}J^{f}_{pV1}+|V_{f}|^{2}J^{f}_{pV2}\right)+|\hat{A_{\chi}}|^{2}\left(m_{Z}^{2}-4m_{\chi}^{2}\right)\left(|A_{f}|^{2}J^{f}_{pA1}+|V_{f}|^{2}J^{f}_{pA2}\right)\right]~,
\eeq
involving the following factors which depend on the various masses
\beq
J^{f}_{pV1} &= 2m_{Z}^{4}\left(m_{f}^{2}-m_{\chi}^{2}\right)\left(-2m_{\chi}^{2}\left(46m_{f}^{2}+m_{Z}^{2}\right)+11m_{f}^{2}m_{Z}^{2}+56m_{\chi}^{4}\right)\\[5pt]
J^{f}_{pV2}& = m_{Z}^{4}\left(-11 m_f^4 m_{Z}^2+4 m_\chi^4 \left(14 m_f^2+m_{Z}^2\right)-2 m_f^2 m_\chi^2 \left(m_{Z}^2-46 m_f^2\right)-112 m_\chi^6\right)\\[5pt]
J^{f}_{pA1}&= -23 m_f^4 m_{Z}^4+192 m_f^2 m_\chi^6 + 4 m_f^2 m_\chi^2 m_{Z}^2 \left(30 m_f^2+7m_{Z}^2\right)-8 m_\chi^4 \left(30 m_f^4+12 m_f^2 m_{Z}^2+m_{Z}^4\right)\\[5pt]
J^{f}_{pA2}&=4m_{Z}^{4}\left(m_{f}^{4}+m_{f}^{2}m_{\chi}^{2}-2m_{\chi}^{4}\right)~.
\eeq

For $m_{\chi}>m_{W^{+},W^{-}}$ DM can also annihilate to  $W$ boson pairs and the thermally averaged DM annihilation cross-section of this process is given by
\beq\label{eq:fDMannZ2}
\langle \sigma v \rangle_{\chi\bar{\chi}\to W^{+}W^{-}} = R_{0}\left(R_{s}+v^{2}R_{p}\right)~.
\eeq
Similar to the fermion channel we parameterise this piece-wise, with an overall factor 
\beq
R_{0} = \frac{\pi \alpha_{\rm em}\kappa^{2}g^{2}}{192\pi\tan\theta_W \cos^2\theta_W}\frac{\sqrt{m_{\chi}^{2}-m_{W}^{2}}}{m_{W}^{4}m_{\chi}\left(m_{Z}^{2}-4m_{\chi}^{2}\right)^{3}}~,
\eeq
where where $\alpha_{\rm em}$ is the fine-structure constant. The `$s$-wave' factor $R_{s}$ is given by
\beq
R_{s} =12|\hat{V}_{\chi}|^{2} \left(m_{Z}^{2}-4m_{\chi}^{2}\right) \left(-3 m_W^6-17
   m_W^4 m_\chi^2+16 m_W^2 m_\chi^4+4
   m_\chi^6\right)~,
\eeq
and the $p$-wave factor can be written as
\beq
R_{p} = \left(|\hat{A}_{\chi}|^{2}P_{A}+|\hat{V}_{\chi}|^{2}P_{V}\right)~,
\eeq
in terms of the subfactors 
\beq
P_{A}& =4\left(-3 m_W^6-17 m_W^4 m_\chi^2+16 m_W^2
   m_\chi^4+4 m_\chi^6\right) \left(m_Z^2-4
   m_\chi^2\right) \\
  P_{V}& = \left(P_{V6}m_{W}^{6}+P_{V4}m_{W}^{4}+P_{V2}m_{W}^{2}+P_{V0}\right) ~,
\eeq
with
\beq
P_{V6} &= 33m_{Z}^{2}-256m_{\chi}^{2},\quad\quad\quad\quad
P_{V4} = 94m_{\chi}^{2}m_{Z}^{2}-1192m_{\chi}^{4}\\[5pt]
P_{V2} & = 464m_{\chi}^{6}+76m_{\chi}^{4}m_{Z}^{2}
\hspace{12mm} P_{V0}  = 40m_{\chi}^{6}m_{Z}^{2}+32m_{\chi}^{8}~.
\eeq

For $m_{\chi}>m_{Z}$ there is also an annihilation channel is to $Z$-pairs and the thermally averaged cross-section for which can be expressed as
\beq\label{eq:fDMannZ3}
&\langle \sigma v \rangle_{\chi\bar{\chi}\to ZZ}= T_{0}\left[T_{s}-v^{2}T_{p}\right]~,
\eeq
with an overall factor,
\begin{align}
T_{0} = \frac{g^{2}\kappa^{2}}{384\pi\cos^{2}\theta_{W}}\frac{\sqrt{m_{\chi}^{2}-m_{Z}^{2}}}{m_{\chi}\left(m_{Z}^{3}-2m_{\chi}^{2}m_{Z}\right)^{4}}~,
\end{align}
with an $s$-wave factor
\beq
T_{s} = 12\left(m_{\chi}^{2}-m_{Z}^{2}\right)\left(m_{Z}^{3}-2m_{\chi}^{2}m_{Z}\right)^{2}\left(|\hat{A}_\chi|^4 m_{Z}^2+2 |\hat{A}_\chi|^2 |\hat{V}_\chi|^2 \left(4 m_{\chi}^{2}-3
   m_{Z}^2\right)+m_{Z}^2 |\hat{V}_\chi|^4\right)~,
\eeq
 and a $p$-wave factor
 \beq
 T_{p} = \left(|\hat{A}_{\chi}|^{4}K_{A}+|\hat{A}_{\chi}|^{2}|\hat{V}_{\chi}|^{2}K_{AV}+|\hat{V}_{\chi}|^{4}K_{V}\right)~,
 \eeq
 where the subfactors are defined as
 \beq
  K_{A} &= \left(128 m_\chi^{10}+23 m_{Z}^{10}-118 m_\chi^2 m_{Z}^8+172
   m_\chi^4 m_{Z}^6+32 m_\chi^6 m_{Z}^4-192 m_\chi^8 m_{Z}^2\right)\\[5pt]
   K_{AV} & = -2m_{Z}^{2}  \left(160 m_\chi^8+21 m_{Z}^8-182 m_\chi^2 m_{Z}^6+508 m_\chi^4 m_{Z}^4-528 m_\chi^6 m_{Z}^2\right)\\[5pt]
   K_{V} & = m_{Z}^{6} \left(76 m_\chi^4+23 m_{Z}^4-66 m_\chi^2 m_{Z}^2\right)~,
 \eeq
\noindent
Finally, for the large values of the DM mass there is also a the contribution from $Zh$ channel. The corresponding cross-section is given by:
\beq
\label{eq:fDMannZ4}
 \langle \sigma v \rangle_{\chi\bar{\chi}\to  Zh} =F_0\left[ F_s
 -\frac{v^2   F_p }{\Big(m_Z^2-4 m_\chi^2\Big)^3
   \Big((m_h-m_Z)^2-4 m_\chi^2\Big) \Big((m_h+m_Z)^2-4 m_\chi^2\Big)} \right]~,
\eeq
with an overall factor 
\beq
F_0=\frac{4 m_{Z}^{4}}{v_{0}^{2}}\frac{g^{2}\kappa^{2}}{16 \cos^2\theta_W}
\frac{\sqrt{m_h^4-2 m_h^2 \Big(4 m_\chi^2+m_Z^2\Big)+\Big(m_Z^2-4 m_\chi^2\Big)^2}}{3072 \pi  m_\chi^4 m_Z^6} ~,
\eeq
an $s$-wave factor 
\beq
F_s=~&3 |\hat{A}_\chi|^2 \Big(m_h^4-2 m_h^2 \Big(4 m_\chi^2
+m_Z^2\Big)+\Big(m_Z^2-4 m_\chi^2\Big)^2\Big)
\\ \qquad\qquad
 & +\frac{3 m_Z^4 |\hat{V}_\chi|^2 \Big(-8 m_\chi^2 \Big(m_h^2-5
   m_Z^2\Big)+\Big(m_h^2-m_Z^2\Big)^2+16 m_\chi^4\Big)}{\Big(m_Z^2-4 m_\chi^2\Big)^2}~,
   \eeq
and a $p$-wave factor
   \beq
F_p=&	|\hat{A}_\chi|^2 \Big(m_Z^2-4 m_\chi^2\Big) \Big(G_A \times G_A' \Big) + m_Z^4 |\hat{V}_\chi |^2 G_V~,
\eeq
with the subfactors
\beq
G_A&=96 m_\chi^6
 (5 m_h^2+7 m_Z^2)+5 m_Z^4 (m_h^2-m_Z^2)^2+8 m_\chi^4 (12 m_h^4+6 m_h^2 m_Z^2+43 m_Z^4)\\[5pt]
   &\quad\quad 
     -2 m_\chi^2 m_Z^2 (24m_h^4-37 m_h^2 m_Z^2+59 m_Z^4)
   +384 m_\chi^8~,
   \\[8pt]
G_A'&=m_h^4-2 m_h^2 \Big(4 m_\chi^2+m_Z^2\Big)+\Big(m_Z^2-4 m_\chi^2\Big)^2\\[8pt]
G_V&=128 m_\chi^8 (37 m_h^2-82 m_Z^2)+5 m_Z^2 (m_h^2-m_Z^2)^4-3584 m_\chi^{10}\\[5pt]
&\quad\quad  +32 m_\chi^6 (-69 m_h^4+217 m_h^2 m_Z^2+242 m_Z^4)~,
   \\[8pt]
&\quad \quad   +2 m_\chi^2 (m_h^2-m_Z^2)^2 (-16 m_h^4+m_h^2 m_Z^2+37 m_Z^4)\\[5pt]
&\quad\quad   +8 m_\chi^4 (55
   m_h^6-178 m_h^4 m_Z^2+147 m_h^2 m_Z^4-200 m_Z^6) ~.
\eeq

\section{Dark matter production by $\phi$ field decay through Standard Model loops }
\label{DMdensityPhiDecay}

Dark matter  production due to $\phi$ decay will inherently occur during the era of significant $\phi$ decays and as such one must treat the evolution of the abundances with care. First we will examine the Boltzmann equations and discuss the temperature evolution of the thermal bath during this period, following \cite{Chung:1998rq, Giudice:2000ex} and then we give the loop induced branching fraction of $\phi$ to DM for the Higgs portal with scalar DM and fermion  DM.

\subsection{Dark matter production due to $\phi$ decay}

The Boltzmann equations which describes the evolution of the decaying state $\phi$, the DM  $\chi$, and the radiation $R$ are given by
     \beq\label{eq:phiDecay1}
     &\frac{d\rho_{\phi}}{dt} = -3H\rho_{\phi} - \Gamma\rho_{\phi}\\
     &\frac{d\rho_{R}}{dt} = -3H\rho_{R} + \Gamma\rho_{\phi}+\langle\sigma v\rangle 2\langle E_{\chi}\rangle \left[n_{\chi}^{2}-\left(n_{\chi}^{{\rm eq}}\right)^{2}\right]\\
     & \frac{dn_{\chi}}{dt} = -3Hn_{\chi} - \langle\sigma v\rangle \left[n_{\chi}^{2}-\left(n_{\chi}^{{\rm eq}}\right)^{2}\right]~,
     \eeq
where the $\phi$ particles decay with a rate $\Gamma$ into radiation, $\chi$ particles get created and annihilated into the radiation with a thermally averaged cross-section $\langle\sigma v\rangle$, with $ 2\langle E_{\chi}\rangle$ average energy released in  the annihilation of the $\chi$ particles and $ \langle E_{\chi}\rangle\simeq \sqrt{M^{2}+9T^{2}}$ being the energy carried by each of the $\chi$ particles.

It is assumed that the $\phi$ dominates the energy density of the universe during the early matter dominated period,\footnote{One can generalise this analysis to the case that the early universe has some arbitrary equation of state (i.e.~other matter domination), see \cite{Maldonado:2019qmp}.} and thus the  starting $\phi$ energy density of the $\phi$ field sets the initial Hubble rate $H_I$ with
\beq\label{eq:phiDecay6}
&\rho_{\phi}|_{I} = \frac{3}{8\pi}M^{2}_{{\rm Pl}}H^{2}_{I}~.
\eeq
Since the $\phi$ decays in eq.~(\ref{eq:phiDecay1}) inject entropy  into the bath it is useful to rewrite the Boltzmann equations in terms of the following dimesionless variables  (taking $a_{I} = T_{{\rm RH}}^{-1}$)
\beq\label{eq:phiDecay3}
& \phi \equiv \rho_{\phi} a_{I}a^{3}; \qquad  R\equiv \rho_{R}a^{4};  \qquad X = n_{\chi} a^{3} ;  \qquad A = a/a_{I}~,
\eeq
with initial conditions $R_{I} = X_{I} = 0;~ A_{I} = 1$ and
$\phi_{I} = (\rho_{\phi} a^{4}A^{3})|_{I}  = \frac{3}{8\pi}M^{2}_{{\rm Pl}}H^{2}_{I}/T_{{\rm RH}}^{4}$.

Following the derivation in \cite{Giudice:2000ex}, it can be shown that the temperature  of the system evolves according to
\beq\label{eq:phiDecay11}
T = T_{{\rm Max}}f(A)~,
\eeq
where
\beq
 T_{{\rm Max}}\simeq\left(\frac{3}{8}\right)^{2/5}\left(\frac{5}{\pi^{3}}\right)^{1/8}\frac{g_{\ast}^{1/8}(T_{{\rm RH}})}{g_{\ast}^{1/4}(T_{{\rm Max}})}M_{{\rm Pl}}^{1/4}H_{I}^{1/4}T_{{\rm RH}}^{1/2}~,
\eeq
and 
\beq
 f(A) =\xi\frac{(A^{5/2}-1)^{1/4}}{A} = \xi\left(A^{-3/2}-A^{-4}\right)^{1/4}~,
 \eeq
with $\xi\sim \mathcal{O}(1)$ factor. This function $f(A)$ increases from  0 to 1 at  $A=A_{0} \equiv \left(\frac{8}{3}\right)^{2/5}$ corresponding to $T = T_{{\rm Max}}$, and then decreases as $A^{-3/8}$. Thus for $A>A_{0}$ the temperature evolves according to
\beq \label{eq:phiDecay12}
 T &\simeq  T_{{\rm Max}} A^{-3/8}~.
\eeq

As the thermal bath cools from a temperature $T_{\rm Max}$ to $T_{\rm RH}$ the scale factor relation is given by $T\propto a^{-3/8}$. In the interval between the point when $T=T_{{\rm Max}}$ and the $H=\Gamma$  one has $\phi\simeq\phi_{I}$ and it follows from \eqref{eq:phiDecay12} that the Hubble rate can be expressed as \cite{Giudice:2000ex} 
\beq\label{eq:phiDecay14}
H &= \left[\frac{5\pi^{3}g_{\ast}^{2}(T)}{9g_{\ast}(T_{{\rm RH}})}\right]^{1/2}\frac{T^{4}}{T_{{\rm RH}}^{2}M_{{\rm Pl}}}~.
\eeq

Having discussed the evolution of the thermal bath, we next consider the loop induced production of DM due to $\phi$ decays, following \cite{Kaneta:2019zgw}. As discussed in the text we assume the direct coupling of $\phi$ to DM is small or absent, the converse case in which direct production is important has been studied in e.g.~\cite{Gelmini:2006pw}.
 Assuming that at early time the DM number density is small relative to the thermal bath $n_{\chi}^{2}\ll n_{R }^{2}$ then the Boltzmann equation describing time evolution of  $n_{\chi}$ is \cite{Kaneta:2019zgw}
\beq\label{eq:DMnumberDensity1}
&\frac{dn_{\chi}}{dt} + 3Hn_{\chi} = \mathcal{R}(T)~,
\eeq
where   $\mathcal{R}(T)$ is an interaction rate which is a function of the branching fraction to DM $\mathcal{B}_{\rm DM}$, the $\phi$ field energy density, mass, and the decay rate. Following  \cite{Kaneta:2019zgw}  we take 
\beq\label{eq:DMabundance2}
 \mathcal{R}_{{\rm decay}}(T) = 2 \Gamma\frac{\rho_{\phi}}{m_{\phi}}\mathcal{B}_{\rm DM}
\simeq 2\sqrt{\frac{5\pi^{7}}{1296}}\frac{g_{\ast}^{2}(T)}{g_{\ast}^{1/2}(T_{{\rm RH}})}\frac{T_{{\rm RH}}^{-2}}{M_{{\rm Pl}}}\frac{\mathcal{B}_{\rm DM}}{m_{\phi}}T^{8}~.
\eeq
Furthermore, we use the following definition of the DM comoving yield \cite{Kaneta:2019zgw}
\beq\label{eq:DMnumberDensity2}
Y = \frac{n_{\chi}}{T^{8}}~.
\eeq
This form is appropriate since $Y = \frac{n}{s}\propto \frac{n}{a^{3}} \propto \frac{n}{T^{8}}$.
It follows that the  evolution of $Y$ can be expressed in the following manner
\beq\label{eq:DMnumberDensity3}
 \frac{dY}{dT} = -\frac{8}{3}\frac{\mathcal{R}(T)}{HT^{9}}
\simeq -\frac{4\pi^{2}}{9}g_{\ast}(T)\frac{\mathcal{B}_{\rm DM}}{m_{\phi}T^{5}}~,
\eeq
where we have substituted for $H$ and $\mathcal{R}(T)$ using eqns.~(\ref{eq:phiDecay14}) \& (\ref{eq:DMabundance2})  in the latter equality. Then integrating from $T_{{\rm Max}}$ and $T_{{\rm RH}}$ one obtains a yield at $T=T_{\rm RH}$ of the form
\beq\label{eq:DMabundance4}
Y(T_{{\rm RH}}) &\simeq \frac{\pi^{2}}{9}g_{\ast}(T_{{\rm RH}})\frac{\mathcal{B}_{\rm DM}}{m_{\phi}T_{{\rm RH}}^{4}}~.
\eeq

Since we assume that entropy in the bath is conserved after $\phi$ decays are completed, the temperature subsequently evolves with the scale factor as $T\propto a^{-1}$ and thus
\beq\label{eq:DMabundance6}
n_\chi(T_{0}) &= \frac{T_{0}^{3}}{T_{{\rm RH}}^{3}}\frac{g_{\ast}(T_0)}{g_{\ast}(T_{\rm RH})}n_\chi(T_{{\rm RH}})~,
\eeq
where $T_0$ is the temperature today.
Therefore the produced  abundance of the DM from the decay of the $\phi$ field has the following form \cite{Kaneta:2019zgw} 
\beq \label{eq:DMabundance7}
\Omega_{\chi, {\rm decay}}h^{2} &\simeq \frac{16\pi}{3}\frac{n_\chi(T_{0})m_{\chi}}{H_{0}^{2}M_{{\rm Pl}}^{2}}
 \simeq1\times 10^{6}~{\rm GeV}^{-1}\left(g_{\ast}(T_{{\rm RH}})\frac{m_{\chi} n_\chi(T_{{\rm RH}})}{T_{{\rm RH}}^{3}} \frac{\mathcal{B}_{\rm DM}T_{{\rm RH}}^{4}}{m_{\phi}}\right)~.
\eeq
Thus parametrically one finds that the DM abundance produced via loop induced $\phi$ decays scales as follows
\beq
\Omega_{\chi,{\rm decay}}h^{2} &\simeq 0.01\times \left(\frac{\mathcal{B}_{\rm DM}}{10^{-8}}\right)\left(\frac{10~{\rm TeV}}{m_{\phi}}\right)
\left(\frac{T_{{\rm RH}}}{1~{\rm GeV}}\right)\left(\frac{m_{\chi}}{15~{\rm GeV}}\right)~.
\eeq

\subsection{Branching ratio $\phi$ field to scalar DM decay through the Higgs loop}

Even if the $\phi$ field has no interaction with the DM, it could produce DM by decaying through Higgs loop. Recall from  eq.~(\ref{Hlag}) the relevant parts of the scalar Higgs portal Lagrangian and we also include a trilinear coupling of the $\phi$ field to Higgs as follows
\beq
\mathcal{L} \supset -\frac{1}{2}\mu_\phi\phi h^{2}-\frac{1}{4}\kappa\chi^{2}h^{2}-y_{f} h\bar{f}f~.
\eeq 
We assume that there is no direct coupling of the DM to $\phi$, however, even in the absence of a direct coupling decays of $\phi$ can produce DM due to loop level interactions involving the mixed quartic $\chi^{2}h^{2}$. Thus one can define the loop induced decay rate of $\phi$ to DM $\Gamma_{\rm loop}$, which we find to be 
\beq
\Gamma_{\rm loop}= \frac{\langle|\mathcal{M}|^{2}\rangle}{32\pi m_{\phi}}\left(1-\frac{4m_{\chi}^{2}}{m_{\phi}^{2}}\right)^{1/2} \simeq \frac{\mu_{\phi}^{2}\kappa^{2}}{8192\pi^{5}m_{\phi}}\sqrt{1-\frac{4m_{\chi}^{2}}{m_{\phi}^{2}}} ~,
\eeq
where $m_\phi$ is the mass of $\phi$.
Furthermore, in the case of no direct $\phi$-DM coupling, one can define the branching ratio of $\phi$ to DM as follows
\beq
\label{brp}
\mathcal{B}_{\rm DM}& \simeq \frac{\Gamma_{{\rm loop}}}{\Gamma_{h}+\Gamma_{{\rm loop}}} ~,
\eeq
where $\Gamma_{h}$ is the tree level decay rates of $\phi$ to the Standard Model Higgs boson. Decays to Higgs bosons is the only direct decay route and since $\phi$ is heavy ($m_t\ll m_\phi$) we can approximate the branching fraction by just considering this route in the decay products to Standard Model states. The partial decay rate of $\phi$ to Higgs pairs is given by
\beq
\Gamma_{h} = \frac{\mu_{\phi}^{2}}{32\pi m_{\phi}}\sqrt{1-\frac{4m_{h}^{2}}{m_{\phi}^{2}}} ~.
\eeq
Putting these expressions together it follows that
\beq
\mathcal{B}_{\rm DM} & \simeq \left[1+\frac{256\pi^{4}}{\kappa^{2}}\left(1-\frac{4m_{h}^{2}}{m_{\phi}^{2}}\right)^{1/2}
\left(1-\frac{4m_{\chi}^{2}}{m_{\phi}^{2}}\right)^{-1/2}\right]^{-1}~,
\eeq
This expression simplifies for the mass hierarchy relevant to the discussion in the main body of this paper. Thus we assume
$m_\chi, m_h\ll m_\phi$ and this simplifies to
\beq
\mathcal{B}_{\rm DM}& 
\sim \frac{\kappa^{2}}{256\pi^{4}}\simeq 4\times10^{-9} \left(\frac{\kappa}{0.01}\right)^2 ~.
\eeq

\subsection{Branching ratio $\phi$ field to fermion DM decay through the Higgs loop}

Next we consider the case of loop induced DM production for the Higgs portal to fermion DM in the EFT description. We assume a similar Lagrangian involving $\phi$ to the case above 
\beq
\mathcal{L} \to -\frac{1}{2}\mu_{\phi}\phi h^{2}-\frac{1}{2\Lambda} h^{2}\bar{\chi}\chi -y_{f}hf\bar{f}~.
\eeq 
Similarly to before we  derive the loop induced $\phi$ partial decay width to DM
\beq
\Gamma^{{\rm loop}}_{\phi\to\bar{\chi}\chi}& = \frac{\langle|\mathcal{M}|^{2}\rangle}{16\pi m_{\phi}}\left(1-\frac{4m_{\chi}^{2}}{m_{\phi}^{2}}\right)^{1/2} \sim \frac{\mu_{\phi}^{2}m_{\phi}}{2048\pi^{5}\Lambda^2}\left(1-\frac{4m_{\chi}^{2}}{m_{\phi}^{2}}\right)^{3/2} ~,
\eeq
and the branching ratio of $\phi$ to DM (as defined in eq.~(\ref{brp}))
\beq
\mathcal{B}_{\rm DM}
&\simeq \left[1+\frac{64\Lambda^{2}\pi^{4}}{m_{\phi}^{2}}\sqrt{1-\frac{4m_{h}^{2}}{m_{\phi}^{2}}}
\left(1-\frac{4m_{\chi}^{2}}{m_{\phi}^{2}}\right)^{-3/2}\right]^{-1}~,
\eeq
Then taking the relevant mass ordering $m_\chi, m_h \ll m_\phi$ this simplifies to
\beq
\mathcal{B}_{\rm DM}
&\simeq \frac{m_{\phi}^{2}}{64\Lambda^{2}\pi^{4}}\sim 10^{-8}\left(\frac{m_\phi}{10^4~{\rm GeV}}\right)^2\left(\frac{10^6~{\rm GeV}}{\Lambda}\right)^2.
\eeq


\end{document}